\documentclass[journal]{IEEEtran}

\usepackage[T1]{fontenc}
\usepackage[utf8]{inputenc}
\usepackage{cite}
\usepackage{amsmath,amssymb,amsfonts,bm}
\usepackage{amsthm}
\usepackage{graphicx}
\usepackage{booktabs}
\usepackage{array}
\usepackage{url}
\usepackage{stfloats}       
\usepackage{tikz}
\usetikzlibrary{arrows.meta,positioning,fit,calc}
\usepackage[caption=false,font=footnotesize]{subfig}
\usepackage{comment}

\usepackage[switch]{lineno}
\usepackage[named]{algo}
\usepackage{algpseudocode}
\usepackage{algorithm}
\algnewcommand{\LineComment}[1]{\State \(\triangleright\) #1}
\algtext*{EndFor}
\algtext*{EndIf}
\algtext*{EndWhile}
\algtext*{EndFunction}

\newtheorem{lemma}{Lemma}
\newtheorem{proposition}{Proposition}
\newtheorem{corollary}{Corollary}

\begin{document}

\title{Variational Blind Equalization for Dual-Polarization Free-Space Quantum Communications}

\author{Anders~M.~Buvarp,~\IEEEmembership{Senior Member,~IEEE},
        Stefan~Werner,~\IEEEmembership{Fellow,~IEEE}
\thanks{This work was supported by the Research Council of Norway.}
\thanks{A. M. Buvarp and S. Werner are with the Department of Electronic Systems, Norwegian University of Science and Technology, Trondheim, Norway. E-mail: anders.buvarp@ntnu.no. Stefan Werner is also with the Department of Information and Communications Engineering, Aalto University, 00076, Finland.}}

\maketitle

\begin{abstract}
Dual-polarization free-space quantum communication links require receiver
equalization under polarization mixing, fading, phase variation, and limited
channel supervision. This paper investigates variational blind and semi-blind equalization using
VISTA (Variational Inference for Signal Tracking and Adaptation), a
variational autoencoder-based probabilistic physical parameter estimator. In the semi-blind branch, channel-relevant statistics
are mapped to the underlying $2\times2$ complex polarization channel. We
compare forward KL, reverse KL, Average Jeffreys, Max-KL, and Adaptive
Jeffreys regularization and analyze directional KL asymmetry and the local
effect of the regularization weight $\zeta$.
Over 500 paired Monte Carlo trials, forward KL is nearly insensitive to
$\zeta$, whereas reverse and symmetric objectives exhibit low-error regions
followed by degradation under stronger regularization. At
$\zeta=3\times10^{-5}$, Average Jeffreys reduces median channel NRMSE by
approximately $4.8\%$ relative to forward KL at the same weight. Reduced KL
asymmetry does not necessarily imply lower channel error.
A separate fully blind QPSK receiver estimates a direct equalizer from
received data only. Over 3000 paired trials, the VISTA variants have nearly
identical median aligned block-NMSE, while their lower mean aligned block-NMSE
relative to the CMA-plus-separation baseline is concentrated in difficult,
large-error trials. Direct-$W$ optimization with Adam reproduces most of
this mean reduction.
In the fully blind benchmark, the tested weak regularizers provide no resolved
improvement over unregularized VISTA. 
\end{abstract}

\begin{IEEEkeywords}
Blind equalization, variational autoencoder, channel estimation,
dual-polarization receivers, free-space optical communications,
quantum communications, Kullback--Leibler divergence, Jeffreys divergence,
probabilistic inference.
\end{IEEEkeywords}

\section{Introduction}

Free-space optical (FSO) and satellite quantum communication can extend
quantum-secure links beyond fixed fiber infrastructure, but practical
receiver processing remains difficult under realistic propagation and
calibration conditions. Continuous-variable quantum key distribution
(CV-QKD) is a particularly relevant setting because coherent detection
and heterodyne measurements connect naturally to the signal processing
and optical-communication literature, including dual-polarization receiver
architectures and polarization-drift compensation methods
\cite{Pereira2023,Almeida2025,Guilherme2025,Almeida2025a}. However,
free-space and satellite links introduce fading, pointing fluctuations,
time-varying transmittance, polarization mixing, residual calibration
mismatch, and phase variations that require continuous receiver adaptation
\cite{Costa2024,Muga2024}.

In dual-polarization FSO receivers, atmospheric turbulence, pointing
jitter, and geometric loss can produce correlated fading across the
received optical fields~\cite{Jaiswal2018,Kim2026}. These effects occur
together with polarization mixing and receiver mismatch, so the two
heterodyne branches cannot in general be treated as statistically
independent scalar observations. Instead, the receiver must jointly track
the coupled polarization channels while compensating for time-varying
amplitude, phase, and polarization transformations.

Conventional coherent optical receivers address this problem using
adaptive $2\times2$ complex MIMO equalization. Blind methods such as the
constant-modulus algorithm and related adaptive equalizers can recover
polarization mixing without continuous pilot transmission, but their
performance depends on the signal model, adaptation dynamics, and channel
structure. In rapidly varying free-space links, the receiver must infer the
current channel state while avoiding excessive dependence on labeled or
pilot samples.

Variational autoencoders (VAEs) provide a probabilistic alternative to
deterministic neural equalizers by representing the unknown state through a
latent random variable and learning an approximate posterior distribution
from the received observations. In the conventional VAE formulation, an
encoder parameterizes an approximate posterior
$q_\phi(\mathbf{z}\mid\mathbf{y})$, while a decoder maps latent samples
to the observation or signal domain. Training balances a reconstruction
term against the Kullback--Leibler (KL) divergence
$D_{\mathrm{KL}}(q_\phi\Vert p)$ between the approximate posterior and a
prescribed latent prior. The conventional choice is therefore directional:
the posterior is regularized toward the prior through
$D_{\mathrm{KL}}(q_\phi\Vert p)$ rather than through a symmetric
distributional discrepancy.

This probabilistic structure has motivated the use of VAEs for blind and
semi-blind channel estimation and equalization. Variational receivers can
represent uncertainty in an unknown channel or transmitted signal while
using prior information through the latent-variable model. Blind VAE
equalization has been demonstrated for communication channels without pilot
symbols, and subsequent work has extended variational equalization to
coherent optical and dual-polarization channels, including time-varying
polarization impairments and free-space optical MIMO links
\cite{Caciularu2020,Lauinger2022,Nunez2025,Qin2025}. These developments
establish VAEs as a viable tool for channel inference, but they largely
retain the conventional one-sided KL regularization inherited from the
standard VAE formulation.

For channel estimation, the latent variable must retain information about
the physical channel while remaining sufficiently regularized for stable
inference. Excessive prior conformity can suppress channel-state information,
whereas strongly directional regularization can distort the posterior. We
therefore examine whether symmetric treatment of the forward and reverse KL
divergences is more suitable for probabilistic channel estimation.

This paper investigates variational blind and semi-blind equalization for
dual-polarization free-space quantum communications. In the semi-blind branch,
the VAE is used as a probabilistic physical parameter estimator that maps
channel-relevant statistics from revealed transmitted samples to an estimate
of the complex $2\times2$ polarization channel and its posterior
representation. In the fully blind branch, the receiver uses received data
only to estimate a direct equalizing transformation. Hence, the blind receiver
estimates an equalizer rather than a unique physical channel matrix.

The semi-blind experiments isolate the effect of variational regularization
under controlled channel estimation. Forward KL is compared with reverse KL
and symmetric or asymmetry-aware alternatives while varying the regularization
weight $\zeta$ independently of the estimator architecture. The fully blind
branch separately evaluates received-only direct equalization under the QPSK
phase-permutation ambiguity.

The objective is not to replace conventional dual-polarization equalization
with a generic neural receiver. We examine how the form and strength of
variational regularization affect probabilistic channel inference. The VAE is
therefore evaluated as a physical parameter estimator rather than as a
generative representation model. The reported results are receiver-level
results. Protocol-level CV-QKD security analysis is outside the scope of this
paper.

The main contributions are as follows:\\
\emph{1) Statistics-based variational channel estimation:} We formulate
dual-polarization free-space quantum equalization as a probabilistic
$2\times2$ complex MIMO inference problem. In the semi-blind branch, the VAE
uses channel-relevant cross- and auto-statistics from revealed transmitted
samples and decodes the latent representation directly to the physical
polarization channel matrix.\\
\emph{2) Fully blind direct equalization and identifiability:} We develop a
received-only QPSK receiver with an identity-centered decoder that estimates
a direct equalizing transformation. Under the noiseless population model, we characterize the phase-permutation
ambiguity of linear equalizers that preserve the independent QPSK output
distribution. Hence, the fully blind receiver estimates an
equalizer rather than a unique physical channel matrix.\\
\emph{3) Directional and symmetric variational regularization:} We systematically compare
forward KL, reverse KL, Average Jeffreys, Max-KL, and Adaptive Jeffreys under
a common estimator architecture. We also define the absolute and normalized
KL-asymmetry measures $A_{\mathrm{KL}}$ and $a_{\mathrm{KL}}$ and show that
reduced directional asymmetry does not necessarily imply lower channel
estimation error.\\
\emph{4) Local analysis of regularization strength:} We derive a local
quadratic regularization path, a sufficient condition under which a small
positive $\zeta$ decreases validation error, and an implicit hypergradient
condition for an interior stationary regularization weight. We also derive
exact relations among the symmetric divergence objectives.\\
\emph{5) Paired Monte Carlo evaluation:} We evaluate dense $\zeta$ sweeps
over $500$ paired trials and the fully blind receiver over $3000$ paired
trials. In the blind benchmark, VISTA has lower paired mean aligned
block-NMSE than the CMA-plus-separation baseline, with the difference
concentrated in large-error trials; the matched Direct-$W$ Adam control
reproduces most of this mean reduction. The tested weak regularizers show
no resolved improvement over unregularized VISTA.

The remainder of this paper is organized as follows.
Section~\ref{sec:bg} reviews related work.
Section~\ref{sec:model} introduces the channel model and estimation
statistics.
Sections~\ref{sec:vae}--\ref{sec:training} present the variational
estimator, regularization objectives, and training procedure.
Section~\ref{sec:blind_equalization} presents the fully blind direct
equalizer.
Section~\ref{sec:simulations} defines the simulation benchmark and baselines.
Section~\ref{sec:results} presents the numerical results.
Section~\ref{sec:disc} discusses the results, limitations, and complexity.
Section~\ref{sec:conclusion} concludes the paper.

\noindent \emph{Notation:}
Bold lowercase and uppercase letters denote vectors and matrices.
The real and complex fields are $\mathbb{R}$ and $\mathbb{C}$; the
Hermitian transpose and complex conjugate are $(\cdot)^H$ and
$\overline{(\cdot)}$; the Euclidean and Frobenius norms are
$\lVert\cdot\rVert_2$ and $\lVert\cdot\rVert_F$; and the identity matrix of
size $n$ is $\bm I_n$. At discrete time $t$, $\bm x_t,\bm y_t,\bm n_t\in\mathbb{C}^2$
denote the transmitted, received, and additive-noise vectors; $\bm H\in\mathbb{C}^{2\times2}$
is the effective dual-polarization channel, assumed constant over an
estimation block; and $\bm X$, $\bm Y$ are the corresponding sample matrices.
Estimated quantities are marked with hats. Probability distributions are
denoted by $P$ and $Q$, with densities $p$ and $q$ when needed.
$D_{\mathrm{KL}}(Q\Vert P)$ denotes the KL divergence; we write
$D_f=D_{\mathrm{KL}}(Q\Vert P)$ and $D_r=D_{\mathrm{KL}}(P\Vert Q)$
for the forward and reverse directions. The approximate posterior and prior
of $\bm z$ are $q_\phi(\bm z\mid\cdot)$ and $p(\bm z)$; for Gaussian
models, $\bm\mu_q$, $\bm\Sigma_q$ and $\bm\mu_p$, $\bm\Sigma_p$ are the
posterior and prior mean and covariance. $\mathcal{N}_{\mathrm C}(\bm\mu,\bm\Sigma)$
denotes a proper complex Gaussian. Bachmann--Landau notation is $\mathcal{O}(\cdot)$.
\vspace{-3mm}
\section{Related Work}
\label{sec:bg}

Continuous-variable quantum key distribution (CV-QKD) is a relevant setting
for dual-polarization free-space links because coherent detection and
heterodyne measurements connect directly to optical-communication receiver
processing. Recent studies have investigated free-space and satellite-to-ground
CV-QKD under atmospheric loss, transmission fluctuations, and background noise
\cite{Dequal2021,Zheng2025}. The present work focuses on receiver processing
rather than composable security analysis.

Classical coherent optical receivers commonly use adaptive $2\times2$ complex
MIMO equalization and polarization-control algorithms
\cite{Savory2010,Noe2010,Roudas2010,Johannisson2011}. Blind methods such as
the constant-modulus algorithm can recover polarization mixing without
continuous pilot information, but adapt equalizer coefficients directly
rather than infer a probabilistic channel state. Variational methods instead
represent unknown signal or channel quantities through latent random variables.
Caciularu and Burshtein applied variational inference to unsupervised channel
equalization and decoding \cite{Caciularu2020}; Lauinger \emph{et al.} and Qin
\emph{et al.} developed VAE-based blind equalization for coherent optical and
free-space MIMO systems \cite{Lauinger2022,Qin2025}.

Symmetric KL formulations have also been studied in variational learning.
Pu \emph{et al.} considered symmetric matching of encoder- and decoder-induced
joint data--latent distributions \cite{Pu2017}, while Chen \emph{et al.}
developed a symmetric VAE and examined its connection to adversarial learning
\cite{Chen2018}. In contrast, this work studies directional and symmetric
posterior-to-prior regularization for physical channel estimation. The
semi-blind experiments isolate this regularization problem within a structural
physical-channel estimator, while the fully blind branch separately evaluates
received-only direct equalization under the QPSK ambiguity.

\begin{figure}[t!]
\centering
\begin{tikzpicture}[
    node distance=5mm,
    >=Latex,
    block/.style={
        draw,
        rounded corners=2pt,
        align=center,
        minimum height=8mm,
        text width=0.82\columnwidth,
        inner sep=4pt,
        font=\footnotesize
    },
    arrow/.style={
        ->,
        thick
    }
]

\node[block] (tx) {
    Dual-polarization transmitter\\
    $\bm x_t=[x_{X,t},\,x_{Y,t}]^{T}$
};

\node[block, below=of tx] (channel) {
    Free-space dual-polarization channel\\
    $\bm y_t=\bm H\bm x_t+\bm n_t$\\
    polarization mixing, fading, differential phase, receiver mismatch
};

\node[block, below=of channel] (rx) {
    Coherent dual-polarization receiver\\
    transmitted and received blocks $\bm X$ and $\bm Y$
};

\node[block, below=of rx] (stats) {
    Channel-relevant statistics\\
    $\bm S_{yx}=\bm Y\bm X^{H}$,
    \qquad
    $\bm S_{xx}=\bm X\bm X^{H}$
};

\node[block, below=of stats] (vae) {
    Variational channel estimator\\
    $q_{\phi}(\bm z\vert\bm s)$
    $\rightarrow$
    $\widehat{\bm H}=g_{\theta}(\bm z)$
};

\node[block, below=of vae] (out) {
    Channel estimate and posterior representation\\
    $\widehat{\bm H}$ and latent posterior representation
};

\node[block, below=of out] (dsp) {
    Downstream receiver processing\\
    equalization, calibration, and security-relevant processing
};

\draw[arrow] (tx) -- (channel);
\draw[arrow] (channel) -- (rx);
\draw[arrow] (rx) -- (stats);
\draw[arrow] (stats) -- (vae);
\draw[arrow] (vae) -- (out);
\draw[arrow] (out) -- (dsp);

\end{tikzpicture}
\caption{System model for variational dual-polarization channel estimation.
The semi-blind branch forms cross- and auto-statistics from transmitted and
received sample blocks and uses a variational estimator to infer the effective
$2\times2$ complex polarization channel and its posterior representation.
The fully blind branch uses received data only to estimate a direct equalizer.}
\label{fig:vista_system_model}
\end{figure}
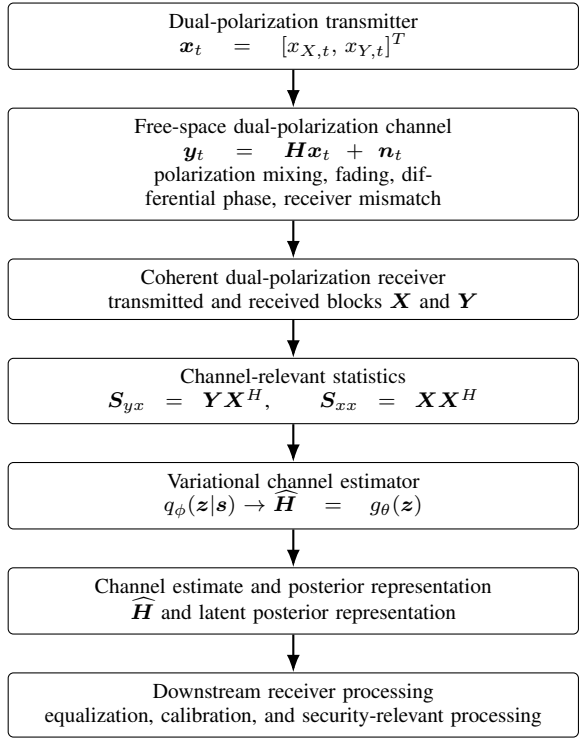
\vspace{-3mm}
\section{Dual-Polarization Free-Space Channel Model}
\label{sec:model}

Fig.~\ref{fig:vista_system_model} summarizes the considered
dual-polarization free-space receiver model and the statistics-based
variational channel estimation procedure. The receiver observes two complex
baseband polarization components whose input--output relation is represented
by a $2\times2$ complex MIMO channel. The objective is to estimate the
effective polarization channel from a finite set of transmitted and received
samples and to obtain an associated posterior representation.
\vspace{-3mm}
\subsection{Dual-Polarization Signal Model}

At symbol time $t$, the transmitted complex baseband vector is
\begin{equation}
    \bm x_t =
    \begin{bmatrix}
        x_{X,t}\\
        x_{Y,t}
    \end{bmatrix}
    \in \mathbb{C}^{2},
\end{equation}
where $x_{X,t}$ and $x_{Y,t}$ denote the symbols transmitted on the two
orthogonal polarization branches. The corresponding received vector is
\begin{equation}
    \bm y_t =
    \bm H \bm x_t + \bm n_t,
    \label{eq:system_model}
\end{equation}
where
$\bm H\in\mathbb{C}^{2\times2}$ denotes the effective dual-polarization
channel and $\bm n_t\in\mathbb{C}^{2}$ denotes the receiver disturbance.

The channel matrix is written as
\begin{equation}
    \bm H =
    \begin{bmatrix}
        h_{XX} & h_{XY}\\
        h_{YX} & h_{YY}
    \end{bmatrix},
\end{equation}
where the diagonal coefficients describe the direct polarization transfer
and the off-diagonal coefficients describe cross-polarization coupling.
The effective channel can incorporate polarization rotation, differential
gain and phase, propagation loss, and residual receiver calibration effects.

For the Gaussian experiments considered in this paper, the receiver
disturbance is modeled as proper complex Gaussian noise,
\begin{equation}
    \bm n_t
    \sim
    \mathcal{N}_\textrm{C}
    \left(
        \bm 0,
        \sigma_n^2 \bm I_2
    \right).
    \label{eq:receiver_noise}
\end{equation}
This model provides a controlled setting for isolating the effect of the
variational regularization objective on channel estimation.

\subsection{Block channel estimation Model}

For a block of $N_p$ transmitted samples, define
\begin{equation}
    \bm X =
    \begin{bmatrix}
        \bm x_1 & \cdots & \bm x_{N_p}
    \end{bmatrix}
    \in\mathbb{C}^{2\times N_p}
\end{equation}
and
\begin{equation}
    \bm Y =
    \begin{bmatrix}
        \bm y_1 & \cdots & \bm y_{N_p}
    \end{bmatrix}
    \in\mathbb{C}^{2\times N_p}.
\end{equation}
The block observation model is therefore
\begin{equation}
    \bm Y = \bm H\bm X + \bm N,
    \label{eq:block_model}
\end{equation}
where $\bm H$ is assumed constant over the estimation block.

Rather than presenting the neural estimator with the raw vectorized
$\bm X$ and $\bm Y$, we form the cross- and auto-statistics
\begin{align}
    \bm S_{yx} &= \bm Y\bm X^{H},\\
    \bm S_{xx} &= \bm X\bm X^{H}.
    \label{eq:channel_statistics}
\end{align}
These quantities retain the second-order structure directly relevant to
linear channel estimation while providing a compact input representation
for the variational estimator.

For reference, the corresponding least-squares channel estimate is
\begin{equation}
    \widehat{\bm H}_{\mathrm{LS}}
    =
    \bm Y\bm X^{H}
    \left(
        \bm X\bm X^{H}
    \right)^{-1},
    \label{eq:ls_channel_estimate}
\end{equation}
when $\bm X\bm X^{H}$ is nonsingular. The proposed variational estimator
does not use $\widehat{\bm H}_{\mathrm{LS}}$ as its input. Instead, it learns
a probabilistic mapping from these statistics to the physical channel
parameters.

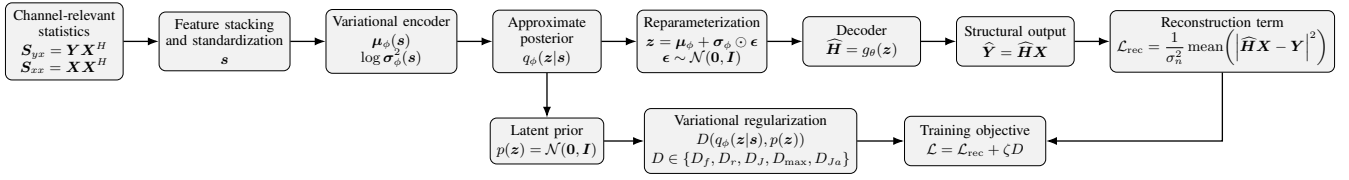
\begin{figure*}[t!]
\centering
\resizebox{0.97\textwidth}{!}{%
\begin{tikzpicture}[
    >=Latex,
    node distance=0.75cm and 0.75cm,
    block/.style={
        draw,
        rounded corners,
        align=center,
        minimum width=2.5cm,
        minimum height=1.05cm,
        fill=gray!10,
        font=\small,
        inner sep=4pt
    },
    smallblock/.style={
        draw,
        rounded corners,
        align=center,
        minimum width=2.3cm,
        minimum height=0.95cm,
        fill=gray!10,
        font=\small,
        inner sep=4pt
    },
    lossblock/.style={
        draw,
        rounded corners,
        align=center,
        minimum width=3.0cm,
        minimum height=1.05cm,
        fill=gray!10,
        font=\small,
        inner sep=4pt
    },
    arrow/.style={->,thick}
]

\node[block] (stats)
{Channel-relevant\\statistics\\[0.5mm]
$\bm S_{yx}=\bm Y\bm X^{H}$\\
$\bm S_{xx}=\bm X\bm X^{H}$};

\node[block, right=of stats] (features)
{Feature stacking\\and standardization\\[0.5mm]
$\bm s$};

\node[block, right=of features] (encoder)
{Variational encoder\\[0.5mm]
$\bm\mu_{\phi}(\bm s)$\\
$\log\bm\sigma_{\phi}^{2}(\bm s)$};

\node[smallblock, right=of encoder] (posterior)
{Approximate\\posterior\\[0.5mm]
$q_{\phi}(\bm z\vert\bm s)$};

\node[block, right=of posterior] (reparam)
{Reparameterization\\[0.5mm]
$\bm z=\bm\mu_{\phi}
+\bm\sigma_{\phi}\odot\bm\epsilon$\\
$\bm\epsilon\sim\mathcal N(\bm0,\bm I)$};

\node[block, right=of reparam] (decoder)
{Decoder\\[0.5mm]
$\widehat{\bm H}=g_{\theta}(\bm z)$};

\node[block, right=of decoder] (predict)
{Structural output\\[0.5mm]
$\widehat{\bm Y}
=\widehat{\bm H}\bm X$};

\node[lossblock, right=of predict] (rec)
{Reconstruction term\\[0.5mm]
$\displaystyle
\mathcal L_{\mathrm{rec}}
=
\frac{1}{\sigma_n^2}
\operatorname{mean}
\!\left(
\left|
\widehat{\bm H}\bm X-\bm Y
\right|^2
\right)$};

\node[smallblock, below=0.95cm of posterior] (prior)
{Latent prior\\[0.5mm]
$p(\bm z)=\mathcal N(\bm0,\bm I)$};

\node[lossblock, right=0.85cm of prior] (div)
{Variational regularization\\[0.5mm]
$D\!\left(q_{\phi}(\bm z\vert\bm s),p(\bm z)\right)$\\[0.5mm]
$D\in\{D_f,D_r,D_J,D_{\max},D_{Ja}\}$};

\node[lossblock, right=1.0cm of div] (objective)
{Training objective\\[0.5mm]
$\mathcal L
=
\mathcal L_{\mathrm{rec}}
+\zeta D$};

\draw[arrow] (stats) -- (features);
\draw[arrow] (features) -- (encoder);
\draw[arrow] (encoder) -- (posterior);
\draw[arrow] (posterior) -- (reparam);
\draw[arrow] (reparam) -- (decoder);
\draw[arrow] (decoder) -- (predict);
\draw[arrow] (predict) -- (rec);

\draw[arrow] (posterior.south) -- (prior.north);
\draw[arrow] (prior) -- (div);
\draw[arrow] (div) -- (objective);

\draw[arrow] (rec.south) |- (objective.east);

\end{tikzpicture}%
}

\caption{Block diagram of the variational channel estimator. The encoder maps
channel-relevant statistics to an approximate posterior over $\bm z$. The
decoder produces the physical channel estimate $\widehat{\bm H}$, which enters
the structural model $\widehat{\bm Y}=\widehat{\bm H}\bm X$. Training combines
the reconstruction term with the variational penalty $\zeta D$.}
\label{fig:vista_vae}
\end{figure*}

\vspace{-3mm}
\section{Variational Channel Estimator}
\label{sec:vae}

The proposed receiver uses a variational autoencoder as a probabilistic
physical parameter estimator for the dual-polarization channel. Rather than
learning a generative representation of the received signal, the model
estimates the effective channel matrix and the associated latent posterior
representation from channel-relevant second-order statistics.
Fig.~\ref{fig:vista_vae} summarizes the estimator architecture and the
corresponding reconstruction and variational-regularization paths.

For each estimation block, the encoder input is formed from the cross- and
auto-statistics defined in Section~\ref{sec:model}. Their real and imaginary
components are stacked into a real-valued feature vector $\bm s$ that forms
the input to the variational encoder.

The encoder parameterizes a diagonal Gaussian approximate posterior
\begin{equation}
    q_{\phi}(\bm z \mid \bm s)
    =
    \mathcal{N}
    \left(
        \bm \mu_{\phi}(\bm s),
        \operatorname{diag}
        \left(
            \bm \sigma_{\phi}^{2}(\bm s)
        \right)
    \right),
    \label{eq:vae_posterior}
\end{equation}
where $\phi$ denotes the encoder parameters. Latent samples are obtained
through the reparameterization
\begin{equation}
    \bm z
    =
    \bm \mu_{\phi}
    +
    \bm \sigma_{\phi}
    \odot
    \bm \epsilon,
    \qquad
    \bm \epsilon\sim\mathcal{N}(\bm 0,\bm I).
    \label{eq:reparameterization}
\end{equation}

The decoder maps the latent variable directly to an estimate of the
complex channel matrix,
\begin{equation}
    \widehat{\bm H}
    =
    g_{\theta}(\bm z),
    \label{eq:vae_channel_decoder}
\end{equation}
where $\theta$ denotes the decoder parameters. The estimated channel is then
inserted into the physical observation model,
\begin{equation}
    \widehat{\bm Y}
    =
    \widehat{\bm H}\bm X.
    \label{eq:vae_prediction}
\end{equation}

The reconstruction term is evaluated in the observation domain as
\begin{equation}
    \mathcal{L}_{\mathrm{rec}}
    =
    \frac{1}{\sigma_n^2}
    \operatorname{mean}
    \left(
        \left|
            \widehat{\bm H}\bm X-\bm Y
        \right|^2
    \right),
    \label{eq:vae_reconstruction}
\end{equation}
where $\sigma_n^2$ denotes the complex receiver-noise variance. This
normalization expresses the reconstruction error relative to the nominal
noise level. The structural decoder therefore forces the latent
representation to support estimation of a physically meaningful channel
matrix rather than unconstrained reconstruction of the received samples.

The complete variational objective is
\begin{equation}
    \mathcal{L}
    =
    \mathcal{L}_{\mathrm{rec}}
    +
    \zeta D,
    \label{eq:vae_objective}
\end{equation}
where $\zeta\geq0$ controls the strength of variational regularization and
$D$ denotes the selected divergence between the approximate posterior
$q_{\phi}(\bm z\mid\bm s)$ and the latent prior $p(\bm z)$.
The conventional VAE uses the forward Kullback--Leibler divergence
$D_{\mathrm{KL}}(q_{\phi}\Vert p)$. In the following section, we examine
whether alternative directional and symmetric regularizers provide a more
appropriate tradeoff between prior conformity and channel-state information
for physical channel estimation.
\vspace{-2mm}
\section{Variational Regularization Objectives}
\label{sec:divergence}

Let
\begin{equation}
D_f
=
D_{\mathrm{KL}}
\left(
q_{\phi}(\bm z\mid\bm s)
\Vert
p(\bm z)
\right)
\label{eq:forward_kl}
\end{equation}
denote the conventional forward KL divergence, and let
\begin{equation}
D_r
=
D_{\mathrm{KL}}
\left(
p(\bm z)
\Vert
q_{\phi}(\bm z\mid\bm s)
\right)
\label{eq:reverse_kl}
\end{equation}
denote the corresponding reverse KL divergence. The standard VAE objective
uses $D_f$, which regularizes the approximate posterior toward the prior in
a single direction.

A symmetric alternative is obtained from the Jeffreys divergence
\begin{equation}
J
=
D_f+D_r.
\label{eq:jeffreys}
\end{equation}
Because $J$ sums both directional terms, its scale is generally larger than
that of either directional KL alone. We therefore use the Average Jeffreys
objective
\begin{equation}
D_J
=
\frac{1}{2}
\left(
D_f+D_r
\right).
\label{eq:average_jeffreys}
\end{equation}
This scaling permits a more direct comparison with the directional
objectives under a common regularization weight $\zeta$.

To quantify the imbalance between the two KL directions, define the
absolute directional asymmetry
\begin{equation}
A_{\mathrm{KL}}
=
\left|
D_f-D_r
\right|
\label{eq:kl_asymmetry}
\end{equation}
and the normalized asymmetry
\begin{equation}
a_{\mathrm{KL}}
=
\frac{
\left|
D_f-D_r
\right|
}{
D_f+D_r+\epsilon
},
\label{eq:normalized_kl_asymmetry}
\end{equation}
where $\epsilon>0$ is a small numerical constant. The normalized quantity satisfies $0\leq a_{\mathrm{KL}}<1$ for
$\epsilon>0$ and is used as a diagnostic of directional imbalance rather than as a standalone training objective.

We additionally consider the worst-direction objective
\begin{equation}
D_{\max}
=
\max
\left(
D_f,D_r
\right),
\label{eq:max_kl}
\end{equation}
which can equivalently be written as
\begin{equation}
D_{\max}
=
D_J+\frac{1}{2}A_{\mathrm{KL}}.
\label{eq:max_kl_decomposition}
\end{equation}
Thus, Max-KL penalizes the larger of the two directional divergences and can
be interpreted as Average Jeffreys augmented by one half of the absolute
directional asymmetry.

Finally, we consider an asymmetry-adaptive Jeffreys objective,
\begin{equation}
D_{Ja}
=
D_J
\left(
1+a_{\mathrm{KL}}^2
\right),
\label{eq:adaptive_jeffreys}
\end{equation}
referred to as Adaptive Jeffreys. Here, ``adaptive'' means that the
regularization magnitude increases with the measured normalized KL
asymmetry. Since $0\leq a_{\mathrm{KL}}<1$, the multiplicative factor lies between
one and two.

The five investigated regularization objectives are therefore Forward KL,
Reverse KL, Average Jeffreys, Max-KL, and Adaptive Jeffreys. In all cases,
the selected divergence $D$ enters the common training objective
\begin{equation}
\mathcal{L}
=
\mathcal{L}_{\mathrm{rec}}
+
\zeta D.
\label{eq:common_regularized_objective}
\end{equation}
so that the experiments separately examine the effect of divergence form
and regularization strength.

For a full batch of estimation blocks, each diagonal Gaussian divergence is
summed over the eight latent coordinates for each block, and the resulting
per-block divergence is averaged over the batch. Average Jeffreys, Max-KL, and
Adaptive Jeffreys are formed per block before batch averaging. In
$\mathcal L_{\mathrm{rec}}$, $\operatorname{mean}$ averages the squared
complex residual over blocks, both polarizations, and samples within each
block.

\subsection{Local Regularization Analysis}
\label{sec:local_regularization}

This section uses a local model of the regularization path. Let
$w\in\mathbb{R}^{P}$ denote the trainable parameter vector. The signal model
and decoded channel matrices are complex-valued, but the encoder and decoder
use real-valued network weights. Let $w_0$ denote a reference solution, such
as the local unregularized solution, and let $w_p$ denote the
divergence-preferred point in the local model.

\begin{proposition}[Local regularization path]
\label{prop:local_regularization_path}
Consider the following local quadratic surrogate near $w_0$:
\begin{align}
\mathcal{L}_{\mathrm{rec}}(w)
&=
\frac{1}{2}(w-w_0)^T\bm A(w-w_0),\\
D(w)
&=
\frac{1}{2}(w-w_p)^T\bm B(w-w_p),
\end{align}
where $\bm A=\bm A^T\succeq\bm0$ and
$\bm B=\bm B^T\succeq\bm0$. If
$\bm A+\zeta\bm B\succ\bm0$, then the unique minimizer of this quadratic
surrogate is
\begin{equation}
w^\star(\zeta)
=
(\bm A+\zeta\bm B)^{-1}
(\bm A w_0+\zeta\bm B w_p).
\label{eq:local_regularization_path}
\end{equation}
For this surrogate, if $\bm A\succ\bm0$, then, as $\zeta\rightarrow0$,
\begin{equation}
w^\star(\zeta)
=
w_0+
\zeta\bm A^{-1}\bm B(w_p-w_0)
+\mathcal{O}(\zeta^2).
\label{eq:local_regularization_expansion}
\end{equation}
If $\bm B\succ\bm0$, then, for the quadratic surrogate,
\[
\lim_{\zeta\rightarrow\infty}w^\star(\zeta)=w_p.
\]
\end{proposition}
\begin{proof}
The stationarity condition for the quadratic surrogate is
\begin{equation}
\bm A(w-w_0)+\zeta\bm B(w-w_p)=\bm0.
\end{equation}
Solving for $w$ gives
\begin{equation}
w^\star(\zeta)
=
(\bm A+\zeta\bm B)^{-1}
(\bm A w_0+\zeta\bm B w_p).
\end{equation}
If $\bm A\succ\bm0$, expanding
$(\bm A+\zeta\bm B)^{-1}$ about $\zeta=0$ gives
\begin{equation}
w^\star(\zeta)
=
w_0+\zeta\bm A^{-1}\bm B(w_p-w_0)
+\mathcal{O}(\zeta^2).
\end{equation}
If $\bm B\succ\bm0$, divide the expression for
$w^\star(\zeta)$ by $\zeta$ and let $\zeta\rightarrow\infty$.
This gives $w^\star(\zeta)\rightarrow w_p$.
\end{proof}

The small-$\zeta$ expansion describes the local displacement from $w_0$
toward $w_p$, while the $\zeta\rightarrow\infty$ limit is a property only
of the quadratic surrogate and does not track the global nonconvex VAE
solution. The surrogate limit should therefore not be interpreted as a
unique divergence-preferred solution or as a bias--variance decomposition
of the full encoder--decoder network.

\begin{corollary}[Sufficient condition for local validation improvement]
\label{cor:local_validation_improvement}
Assume additionally that $\bm A\succ\bm0$.
Let $\mathcal{E}_{\mathrm{val}}(w)$ be differentiable at $w_0$ and define
$g_v=\nabla_w\mathcal{E}_{\mathrm{val}}(w_0)$. Under the assumptions of
Proposition~\ref{prop:local_regularization_path}, if
\begin{equation}
g_v^T\bm A^{-1}\bm B(w_p-w_0)<0,
\label{eq:local_validation_improvement_condition}
\end{equation}
then there exists $\bar{\zeta}>0$ such that
\begin{equation}
\mathcal{E}_{\mathrm{val}}(w^\star(\zeta))
<
\mathcal{E}_{\mathrm{val}}(w_0),
\qquad
0<\zeta<\bar{\zeta}.
\end{equation}
\end{corollary}
\begin{proof}
From Proposition~\ref{prop:local_regularization_path},
\begin{equation}
\left.
\frac{d w^\star}{d\zeta}
\right|_{\zeta=0}
=
\bm A^{-1}\bm B(w_p-w_0).
\end{equation}
Hence,
\begin{equation}
\left.
\frac{d}{d\zeta}
\mathcal{E}_{\mathrm{val}}(w^\star(\zeta))
\right|_{\zeta=0}
=
g_v^T\bm A^{-1}\bm B(w_p-w_0).
\end{equation}
If this derivative is negative, continuity gives
\[
\mathcal{E}_{\mathrm{val}}(w^\star(\zeta))
<
\mathcal{E}_{\mathrm{val}}(w_0)
\]
for all sufficiently small $\zeta>0$.
\end{proof}

Hence, the condition above is sufficient for a small nonzero regularization
weight to improve local validation error: the induced parameter displacement
must decrease the validation criterion. A zero first-order change does not
exclude improvement at higher order. In the quadratic surrogate, if $\bm B\succ\bm0$ and
$\mathcal{E}_{\mathrm{val}}(w_p)>\mathcal{E}_{\mathrm{val}}(w_0)$, large
regularization pulls the solution toward a validation-inferior point. This
local result is consistent with the observed low-error region followed by
degradation at larger regularization weights.

\begin{proposition}[Local hypergradient condition for the regularization weight]
\label{prop:local_hypergradient}
Let
\begin{equation}
\mathcal{L}_{\mathrm{tr}}(w,\zeta)
=
\mathcal{L}_{\mathrm{rec}}(w)+\zeta D(w),
\end{equation}
and let $w^\star(\zeta)$ be a locally differentiable stationary branch that
satisfies
\begin{equation}
\nabla_w\mathcal{L}_{\mathrm{tr}}(w^\star(\zeta),\zeta)=\bm0.
\label{eq:inner_stationarity}
\end{equation}
Assume that $\mathcal{L}_{\mathrm{tr}}$ is twice continuously differentiable
near $(w^\star(\zeta),\zeta)$ and that
\begin{equation}
\bm H_\zeta
=
\nabla^2_{ww}\mathcal{L}_{\mathrm{tr}}(w^\star(\zeta),\zeta)
\end{equation}
is nonsingular. Then
\begin{equation}
\frac{d w^\star}{d\zeta}
=
-\bm H_\zeta^{-1}\nabla_wD(w^\star).
\label{eq:local_hypergradient_path}
\end{equation}
If $\mathcal{E}_{\mathrm{val}}(w)$ has no explicit dependence on $\zeta$,
then
\begin{equation}
\frac{d}{d\zeta}
\mathcal{E}_{\mathrm{val}}(w^\star(\zeta))
=
-
\nabla_w\mathcal{E}_{\mathrm{val}}(w^\star)^T
\bm H_\zeta^{-1}
\nabla_wD(w^\star).
\label{eq:local_hypergradient}
\end{equation}
Hence, any interior stationary point $\zeta^\star$ of the outer validation
objective satisfies
\begin{equation}
\nabla_w\mathcal{E}_{\mathrm{val}}(w^\star(\zeta^\star))^T
\bm H_{\zeta^\star}^{-1}
\nabla_wD(w^\star(\zeta^\star))
=
0.
\label{eq:local_hypergradient_stationarity}
\end{equation}
\end{proposition}
\begin{proof}
Differentiating \eqref{eq:inner_stationarity} with respect to $\zeta$ gives
\begin{equation}
\bm H_\zeta\frac{dw^\star}{d\zeta}
+\nabla_wD(w^\star)
=\bm0.
\end{equation}
Since $\bm H_\zeta$ is nonsingular, this equation gives
\eqref{eq:local_hypergradient_path}. The chain rule applied to
$\mathcal E_{\mathrm{val}}(w^\star(\zeta))$ gives
\eqref{eq:local_hypergradient}. Setting the outer derivative to zero gives
\eqref{eq:local_hypergradient_stationarity}.
\end{proof}

The condition is local and does not imply global optimality or a closed-form
$\zeta^\star$, since all terms depend on $w^\star(\zeta)$.
These results apply to the stated local surrogate. The full neural objective
may violate the smoothness or nonsingularity assumptions because of
overparameterization, ReLU activations, variance clipping, and Max-KL
switching.

\begin{lemma}[Relations among the symmetric objectives]
\label{lem:divergence_relations}
Let $J=D_f+D_r$, $D_J=J/2$, and
$A_{\mathrm{KL}}=|D_f-D_r|$. Then
\begin{equation}
D_{\max}
=
D_J+\frac{1}{2}A_{\mathrm{KL}},
\end{equation}
and, for $J>0$,
\begin{equation}
D_J
\leq
D_{Ja}
\leq
D_{\max}
\leq
2D_J.
\label{eq:divergence_objective_bounds}
\end{equation}
When $\epsilon=0$,
\begin{equation}
D_{Ja}-D_J
=
\frac{A_{\mathrm{KL}}^2}{2J},
\qquad
D_{\max}-D_J
=
\frac{1}{2}A_{\mathrm{KL}}.
\end{equation}
\end{lemma}
\begin{proof}
For nonnegative $D_f$ and $D_r$,
\begin{equation}
\max(D_f,D_r)
=
\frac{D_f+D_r+|D_f-D_r|}{2},
\end{equation}
which gives
$
D_{\max}=D_J+\frac{1}{2}A_{\mathrm{KL}}.
$
Since $0\leq A_{\mathrm{KL}}\leq J$ and
\[
a_{\mathrm{KL}}
=
\frac{A_{\mathrm{KL}}}{J+\epsilon},
\]
we have
\[
J a_{\mathrm{KL}}^2
=
\frac{J A_{\mathrm{KL}}^2}{(J+\epsilon)^2}
\leq A_{\mathrm{KL}}.
\]
Hence,
\[
D_{Ja}
=
D_J+\frac{1}{2}J a_{\mathrm{KL}}^2
\leq
D_J+\frac{1}{2}A_{\mathrm{KL}}
=
D_{\max}.
\]
Together with $a_{\mathrm{KL}}^2\geq0$ and
$A_{\mathrm{KL}}\leq J$, this gives
\[
D_J\leq D_{Ja}\leq D_{\max}\leq 2D_J.
\]
When $\epsilon=0$, $a_{\mathrm{KL}}=A_{\mathrm{KL}}/J$. Substitution
into $D_{Ja}=D_J(1+a_{\mathrm{KL}}^2)$ gives
\[
D_{Ja}-D_J=\frac{A_{\mathrm{KL}}^2}{2J}.
\]
The second identity follows from $D_{\max}=D_J+\frac{A_{\mathrm{KL}}}{2}$.
\end{proof}


\section{Training and Estimation Procedure}
\label{sec:training}

Training minimizes \eqref{eq:common_regularized_objective} using Forward KL,
Reverse KL, Average Jeffreys, Max-KL, or Adaptive Jeffreys, with
standardized channel-relevant statistics and the same standardization applied
at evaluation time. The weight $\zeta$ controls the regularization strength
independently of the estimator architecture; encoder and decoder parameters
are optimized jointly. Numerical settings are given in
Section~\ref{sec:simulations}.
\vspace{-5mm}
\subsection{Channel Estimate and Evaluation}

At evaluation time, the posterior mean is used as the decoder input,
\begin{equation}
    \widehat{\bm H}
    =
    g_{\theta}\!\left(\bm\mu_{\phi}(\bm s)\right),
    \label{eq:posterior_mean_channel}
\end{equation}
which removes latent-sampling variation from the reported estimate.
Because the decoder is nonlinear, this decoded mean is not generally equal
to the posterior mean $\mathbb{E}_{q_\phi}[g_\theta(\bm z)]$.
The reported posterior variances are therefore treated as latent-space
diagnostics rather than calibrated physical-channel uncertainties.

Channel estimation accuracy is measured by
\begin{equation}
    \operatorname{NRMSE}
    =
    \frac{
        \left\|\widehat{\bm H}-\bm H\right\|_{F}
    }{
        \left\|\bm H\right\|_{F}+\epsilon
    },
    \label{eq:channel_nrmse}
\end{equation}
where $\epsilon>0$ is a small numerical constant.

We additionally record the forward and reverse KL divergences, Average
Jeffreys divergence, $A_{\mathrm{KL}}$, $a_{\mathrm{KL}}$, and the posterior
variance to relate posterior regularization to channel estimation performance.

\section{Fully Blind Direct Equalization}
\label{sec:blind_equalization}

The fully blind receiver observes $Y$ only and estimates the direct
equalizer $W_c$ rather than a unique physical channel matrix.

\subsection{Received-Only Features and Direct Equalizer}

For a received block $\bm Y\in\mathbb{C}^{2\times N_b}$, define
\begin{equation}
\bm S_{yy}
=
\frac{1}{N_b}\bm Y\bm Y^H,
\qquad
p_k[t]=|y_k[t]|^2,
\quad k\in\{1,2\}.
\end{equation}
The receiver uses the following 16 real-valued moment features:
\begin{align}
\bm f_{\mathrm{mom}}
=
\big[
&\operatorname{vec}_{r}(\Re\{\bm S_{yy}\})^T,\,
\operatorname{vec}_{r}(\Im\{\bm S_{yy}\})^T,\,
m_{4,1},\,m_{4,2},\,m_{22}, \nonumber\\
&\Re\{c_4\},\,\Im\{c_4\},\,
c_{\mathrm{CM},1},\,c_{\mathrm{CM},2},\,c_{\mathrm{cov}}
\big]^T,
\label{eq:blind_moment_features}
\end{align}
where $\operatorname{vec}_{r}$ denotes row-major vectorization and
\begin{align}
m_{4,k}
&=
\frac{1}{N_b}\sum_{t=1}^{N_b}p_k[t]^2,\\
m_{22}
&=
\frac{1}{N_b}\sum_{t=1}^{N_b}p_1[t]p_2[t],\\
c_4
&=
\frac{1}{N_b}\sum_{t=1}^{N_b}
y_1[t]^2\overline{y_2[t]}^2,\\
c_{\mathrm{CM},k}
&=
\frac{1}{N_b}\sum_{t=1}^{N_b}(p_k[t]-1)^2,\\
c_{\mathrm{cov}}
&=
\frac{1}{N_b}\sum_{t=1}^{N_b}
(p_1[t]-\overline p_1)(p_2[t]-\overline p_2),\\
\overline p_k
&=
\frac{1}{N_b}\sum_{t=1}^{N_b}p_k[t].
\end{align}

The receiver also uses the first $M=64$ received vectors. This sample slice is
normalized using the received-only RMS:
\begin{equation}
\begin{aligned}
\rho
&=
\left[
\max\left(
\frac{1}{2N_b}
\sum_{k=1}^{2}\sum_{t=1}^{N_b}|y_k[t]|^2,
\epsilon_{\mathrm{mach}}
\right)
\right]^{1/2},\\
\widetilde{\bm Y}
&=
\bm Y/\rho,
\end{aligned}
\end{equation}
where $\epsilon_{\mathrm{mach}}$ denotes machine epsilon. The complete encoder input is
\begin{equation}
\begin{aligned}
\bm v
=
\big[
&\bm f_{\mathrm{mom}}^T,\,
\Re\{\widetilde y_1[1:M]\}^T,\,
\Im\{\widetilde y_1[1:M]\}^T,\\
&\Re\{\widetilde y_2[1:M]\}^T,\,
\Im\{\widetilde y_2[1:M]\}^T
\big]^T
\in\mathbb{R}^{272}.
\end{aligned}
\label{eq:blind_encoder_input}
\end{equation}
This RMS normalization applies only to the received sample slice in
\eqref{eq:blind_encoder_input}. The 16 moment features in
\eqref{eq:blind_moment_features} are computed from the original,
unnormalized $\bm Y$.

The encoder has two fully connected 212-unit ReLU layers. Separate
eight-dimensional heads produce the posterior mean and log-variance:
\begin{equation}
q_\phi(\bm z\mid\bm v)
=
\mathcal{N}\!\left(
\bm\mu_\phi(\bm v),
\operatorname{diag}(\exp\{\bm\ell_\phi(\bm v)\})
\right),
\end{equation}
Each log-variance component is clamped to $[-8,5]$, and
$p(\bm z)=\mathcal{N}(\bm0,\bm I)$. The decoder has two fully connected
212-unit ReLU layers and eight real outputs. For decoder output
$\bm d\in\mathbb{R}^{8}$,
\begin{equation}
\begin{aligned}
\Delta\bm W
&=
\operatorname{reshape}_{2\times2}(\bm d_{1:4})
+\mathrm{j}\operatorname{reshape}_{2\times2}(\bm d_{5:8}),\\
\widehat{\bm W}
&=
\bm I_2+\Delta\bm W.
\end{aligned}
\label{eq:identity_centered_equalizer}
\end{equation}
The reshaping is row-major. The final decoder layer is initialized to zero.
Hence, the initial effective equalizer is $\bm I_2$.
\vspace{-2mm}
\subsection{Blind Objective and Optimization}

The receiver applies the direct equalizer as
\begin{equation}
\widehat{\bm X}
=
\widehat{\bm W}\bm Y.
\end{equation}
For unit-power QPSK, the training loss is
\begin{align}
\mathcal{L}_{\mathrm{CMA}}
&=
\frac{1}{2N_b}
\sum_{k=1}^{2}\sum_{t=1}^{N_b}
\left(|\widehat x_k[t]|^2-1\right)^2,\\
\mathcal{L}_{\mathrm{sep}}
&=
\left|
\frac{1}{N_b}
\sum_{t=1}^{N_b}
\widehat x_1[t]\overline{\widehat x_2[t]}
\right|^2,\\
\mathcal{L}_{\mathrm{blind}}
&=
\mathcal{L}_{\mathrm{CMA}}
+
0.1\mathcal{L}_{\mathrm{sep}}.
\label{eq:blind_objective}
\end{align}
A regularized blind VAE uses
\begin{equation}
\mathcal{L}_{\mathrm{blind}}+\zeta D.
\end{equation}

Each condition trains a new VAE on one received block. Training uses full-block
Adam with learning rate $2\times10^{-3}$ for 1000 epochs. It uses
reparameterized latent samples. Evaluation decodes the posterior mean. Within
each paired trial, all VAE conditions use the same received block, network
initialization, and latent-sampling seed.

The CMA-plus-separation baseline uses the same received block and 1000
adaptation iterations. It starts from $\bm W^{(0)}=\bm I_2$ and uses the batch
update
\begin{align}
\bm Z^{(i)}&=\bm W^{(i)}\bm Y,
\qquad
\bm E^{(i)}=|\bm Z^{(i)}|^2-\bm1,\\
\bm G^{(i)}&=
\frac{(\bm E^{(i)}\odot\bm Z^{(i)})\bm Y^H}{N_b},
\end{align}
The update includes the cross-polarization separation terms
\begin{align}
\bm G^{(i)}_{1,:}
&\leftarrow
\bm G^{(i)}_{1,:}
+0.1c^{(i)}\frac{\bm z^{(i)}_{2,:}\bm Y^H}{N_b},\\
\bm G^{(i)}_{2,:}
&\leftarrow
\bm G^{(i)}_{2,:}
+0.1\overline{c^{(i)}}\frac{\bm z^{(i)}_{1,:}\bm Y^H}{N_b},\\
c^{(i)}
&=
\frac{1}{N_b}\sum_{t=1}^{N_b}
z^{(i)}_1[t]\overline{z^{(i)}_2[t]},\\
\bm W^{(i+1)}
&=
\bm W^{(i)}-10^{-2}\bm G^{(i)}.
\label{eq:cma_baseline_update}
\end{align}

\subsection{QPSK Ambiguity and Evaluation}

Let
\begin{equation}
\mathcal{G}
=
\left\{
\bm P\operatorname{diag}
\left(e^{\mathrm{j}\phi_1},e^{\mathrm{j}\phi_2}\right):
\bm P\in S_2,\;
\phi_1,\phi_2\in
\left\{0,\frac{\pi}{2},\pi,\frac{3\pi}{2}\right\}
\right\}.
\label{eq:qpsk_ambiguity_group}
\end{equation}
Thus, $|\mathcal{G}|=2!4^2=32$. For the received-only carrier-phase
recovery, let
\begin{equation}
u_k[t]=(\widehat{\bm W}\bm Y)_{k,t},
\qquad
\widehat{\alpha}_k
=
\frac{1}{4}
\arg\left(
\frac{1}{N_b}
\sum_{t=1}^{N_b}u_k[t]^4
\right),
\end{equation}
and define the carrier-phase-corrected equalizer as
\begin{equation}
\widehat{\bm W}_{\mathrm{CPR}}
=
\operatorname{diag}
\left(
e^{-\mathrm{j}\widehat{\alpha}_1},
e^{-\mathrm{j}\widehat{\alpha}_2}
\right)
\widehat{\bm W}.
\end{equation}
The evaluation-only aligned block-NMSE is then
\begin{equation}
\operatorname{NMSE}_{\mathcal{G}}
=
\min_{\bm G\in\mathcal{G}}
\frac{
\|\bm G\widehat{\bm W}_{\mathrm{CPR}}\bm Y-\bm X\|_F^2
}{
\|\bm X\|_F^2
}.
\label{eq:aligned_blind_nmse}
\end{equation}
The CMA-plus-separation loss is invariant to arbitrary per-stream phase
rotations. Received-only carrier-phase recovery is therefore applied before
the discrete QPSK phase and permutation alignment used for evaluation.

\begin{proposition}[Noiseless blind QPSK identifiability]
\label{prop:blind_qpsk_identifiability}
Suppose that $\bm Y=\bm H\bm X$, where
$\bm H\in\mathbb{C}^{2\times2}$ is nonsingular and the two source streams are
independent, equiprobable, unit-power QPSK processes. At the population level,
a linear equalizer with the same independent QPSK output distribution is
identifiable only up to the left action of
$\mathcal{G}$:
\begin{equation}
\widehat{\bm W}
\sim
\widetilde{\bm W}
\quad\Longleftrightarrow\quad
\widetilde{\bm W}
=
\bm G\widehat{\bm W},
\qquad
\bm G\in\mathcal{G}.
\end{equation}
Hence, the noiseless equalizer class is $\left\{
\bm G\bm H^{-1}:\bm G\in\mathcal{G}
\right\}$.
The physical channel is ambiguous under the right action
$\bm H\sim\bm H\bm G^{-1}$. Therefore, fully blind operation does not identify
a unique channel matrix.
\end{proposition}
\begin{proof}
Let $\mathcal Q=\{1,\mathrm{j},-1,-\mathrm{j}\}$ and define
$\bm A=\widehat{\bm W}\bm H$. Under the assumed independent two-stream
QPSK output distribution, $\bm A$ is invertible. Consider a row $(a,b)$
of $\bm A$. Its output is
$U=aX_1+bX_2$. Let $T=X_1\overline{X_2}$. Since $X_1$ and $X_2$ are
independent and equiprobable, $T$ is uniform on $\mathcal Q$. Because $U$ is
QPSK,
\begin{equation}
|U|^2
=
|a|^2+|b|^2+2\Re\{a\overline b\,T\}
\end{equation}
must be constant over the four values of $T$. Setting $T=1$ and $T=-1$
gives $\Re\{a\overline b\}=0$. Setting $T=\mathrm{j}$ and
$T=-\mathrm{j}$ gives $\Im\{a\overline b\}=0$. Hence
$a\overline b=0$. Each row therefore contains at most one nonzero entry.

The QPSK output has unit modulus, so any nonzero entry has unit magnitude.
Preserving the QPSK alphabet requires that entry to belong to $\mathcal Q$.
Invertibility forces the two rows to select different input coordinates.
Hence, the transform is a generalized permutation matrix in $\mathcal G$.
This proves the left-action equivalence class. The channel ambiguity follows from
$\bm H\bm X=(\bm H\bm G^{-1})(\bm G\bm X)$.
\end{proof}

Proposition~\ref{prop:blind_qpsk_identifiability} concerns noiseless
source-model identifiability. It does not state that the finite-sample
CMA-plus-separation objective has no spurious stationary points. It also does
not state that the optimization recovers a unique member of the equivalence
class.
\vspace{-2mm}
\section{Simulation Benchmark}
\label{sec:simulations}

\subsection{Simulation Configuration}

The simulation benchmark evaluates channel estimation performance under
controlled dual-polarization channel and receiver conditions generated
according to \eqref{eq:system_model}. Each estimation block contains
$N_p=32$ dual-polarization transmitted samples. The transmitted symbols are
proper complex Gaussian with variance $4$ per complex symbol. For each block,
an independent static $2\times2$ complex channel is generated as
\begin{equation}
\bm H
=
\begin{bmatrix}
\cos\theta &
\sin\theta\,e^{\mathrm{j}\phi}\\
-\sin\theta\,e^{-\mathrm{j}\phi} &
\cos\theta\,e^{\mathrm{j}\psi}
\end{bmatrix},
\label{eq:simulation_channel}
\end{equation}
where
\begin{equation}
\theta\sim\mathcal{U}(0.10,0.70),
\qquad
\phi,\psi\sim\mathcal{U}(-\pi,\pi),
\end{equation}
with the three parameters drawn independently. The receiver disturbance is
proper complex Gaussian with variance $0.02$ per complex component. The benchmark does not sweep \textrm{SNR} as an independent parameter; with the specified signal and disturbance variances, the resulting branch \textrm{SNRs} are approximately $23.0$ dB for the Gaussian experiment and $17.0$ dB for the blind QPSK experiment.

The matrix in~\eqref{eq:simulation_channel} is used as a controlled effective polarization-channel
model rather than as a complete physical propagation model. Its parameters
represent coupled polarization mixing and phase effects, while non-unitary
behavior is permitted to represent effective differential gain or receiver
mismatch.

For each Monte Carlo trial, $1000$ independent training blocks and $1000$
independent test blocks are generated. The encoder input consists of the real
and imaginary parts of
$\bm S_{yx}=\bm Y\bm X^H$ and
$\bm S_{xx}=\bm X\bm X^H$, producing a 16-dimensional real feature vector.
Each feature is standardized using the mean and standard deviation computed
from the training set, with a standard-deviation floor of $10^{-6}$. The same
training-set statistics are applied to the corresponding test features.

The semi-blind encoder contains two fully connected hidden layers of $212$
ReLU units, followed by separate eight-dimensional heads for the posterior
mean and log-variance. The latent dimension is eight, and the decoder contains
two fully connected hidden layers of $212$ ReLU units followed by an
eight-dimensional output representing the real and imaginary components of
the estimated $2\times2$ complex channel matrix. All variational estimators
are trained using full-batch Adam with learning rate $2\times10^{-3}$ for
$1000$ epochs, with no explicit weight initialization rule beyond the seeded
PyTorch initialization.
\vspace{-2mm}
\subsection{Fully Blind QPSK Configuration}

The fully blind benchmark uses one static received block with $N_b=1000$
samples. The two transmitted streams are independent equiprobable unit-power
QPSK symbols from $\{1,\mathrm{j},-1,-\mathrm{j}\}$. The static
dual-polarization Jones channel follows \eqref{eq:simulation_channel}. The
additive disturbance is proper complex Gaussian with variance $0.02$ per
complex component. VISTA and the CMA-plus-separation baseline adapt to the
same received block. 

The Direct-$W$ Adam baseline directly optimizes
\begin{equation}
    \bm W=\bm I_2+\Delta\bm W,
\end{equation}
where $\Delta\bm W$ is parameterized by eight real trainable parameters.
All parameters are initialized to zero, giving
$\bm W^{(0)}=\bm I_2$. The parameters are optimized directly from the
received block using Adam with learning rate $2\times10^{-3}$ for 1000
epochs, with one update per epoch. The objective is the same unregularized
CMA-plus-separation loss component used by VISTA, namely
$\mathcal L_{\mathrm{CMA}}+0.1\mathcal L_{\mathrm{sep}}$.
No transmitted symbols are used during adaptation. Direct-$W$ uses the
same received-only fourth-power carrier-phase recovery and subsequent
QPSK phase/permutation alignment as VISTA for evaluation.
The aligned block-NMSE is then evaluated using
\eqref{eq:aligned_blind_nmse}.

Final receiver adaptation uses $\bm Y$ only. A separate offline development
sweep over candidate $\zeta$ values used $10$ paired trials and simulated
aligned-NMSE with the generated transmitted labels. The resulting operating
points were fixed before the $3000$-trial confirmation:
$\zeta=3\times10^{-6}$ for Forward KL, $\zeta=10^{-7}$ for Reverse KL, and
$\zeta=3\times10^{-6}$ for Average Jeffreys. The development trials were
separate from the confirmation trials and were not included in the reported
statistics. During confirmation, transmitted labels were used only after
adaptation for evaluation.

The fully blind confirmation evaluates Forward KL, Reverse KL, and Average
Jeffreys. Max-KL and Adaptive Jeffreys are evaluated only in the semi-blind
experiments.
\vspace{-2mm}
\subsection{Baselines and Performance Metrics}

Three reference estimators are included. The known-sample least-squares
estimator is $\widehat{\bm H}_{\mathrm{LS}}
=
\bm Y\bm X^H
\left(
\bm X\bm X^H
\right)^{-1}$.
A deterministic neural estimator is trained directly from the same
statistics-based input to the channel target and provides a nonvariational
learned baseline. A zero estimator,
$\widehat{\bm H}=\bm 0$, is included as a trivial reference.

The primary performance measure is the channel NRMSE defined in
\eqref{eq:channel_nrmse}, with $\epsilon=10^{-12}$ in the numerical
experiments. For the variational estimators, we additionally record the
forward and reverse KL divergences, Average Jeffreys divergence, the absolute
directional asymmetry $A_{\mathrm{KL}}$, the normalized asymmetry
$a_{\mathrm{KL}}$, and the posterior variance. These diagnostics are used to
relate posterior regularization to channel estimation accuracy.

For each Monte Carlo trial, the estimator is evaluated on 1000 independent
test blocks. The trial-level channel NRMSE is the arithmetic mean of the
1000 block-level NRMSE values. Reported medians and interquartile ranges are
computed across these trial-level values.

Comparisons among variational objectives use $500$ paired Monte Carlo trials.
Within each trial, all competing objectives use the same channel,
transmitted-signal, and noise realizations. Neural-network initialization
and stochastic latent-sampling streams are also matched across objectives,
thereby reducing variation unrelated to the regularization objective.

The regularization-weight sweep evaluates
\begin{equation}
\begin{aligned}
\zeta \in \{&
10^{-6},\,3\times10^{-6},\,10^{-5},\,3\times10^{-5},\,10^{-4},\\
&3\times10^{-4},\,10^{-3},\,3\times10^{-3},\,10^{-2}
\}.
\end{aligned}
\label{eq:zeta_grid}
\end{equation}
The experiment uses seed $20260817$. Reported curves show the median channel
NRMSE and normalized KL asymmetry across trials, with the interquartile range
used to characterize trial-to-trial variation. Selected paired differences are
additionally evaluated using $10{,}000$ paired nonparametric bootstrap
resamples: records are matched by trial, the bootstrap statistic is the mean
paired difference, and the reported $95$\% confidence interval is the
percentile interval $[2.5, 97.5]$ of the bootstrap distribution.

\begin{figure*}[t!]
    \centering
    \subfloat[Median channel NRMSE versus $\zeta$\label{fig:nrmse_vs_zeta}]
    {\includegraphics[width=0.45\textwidth]{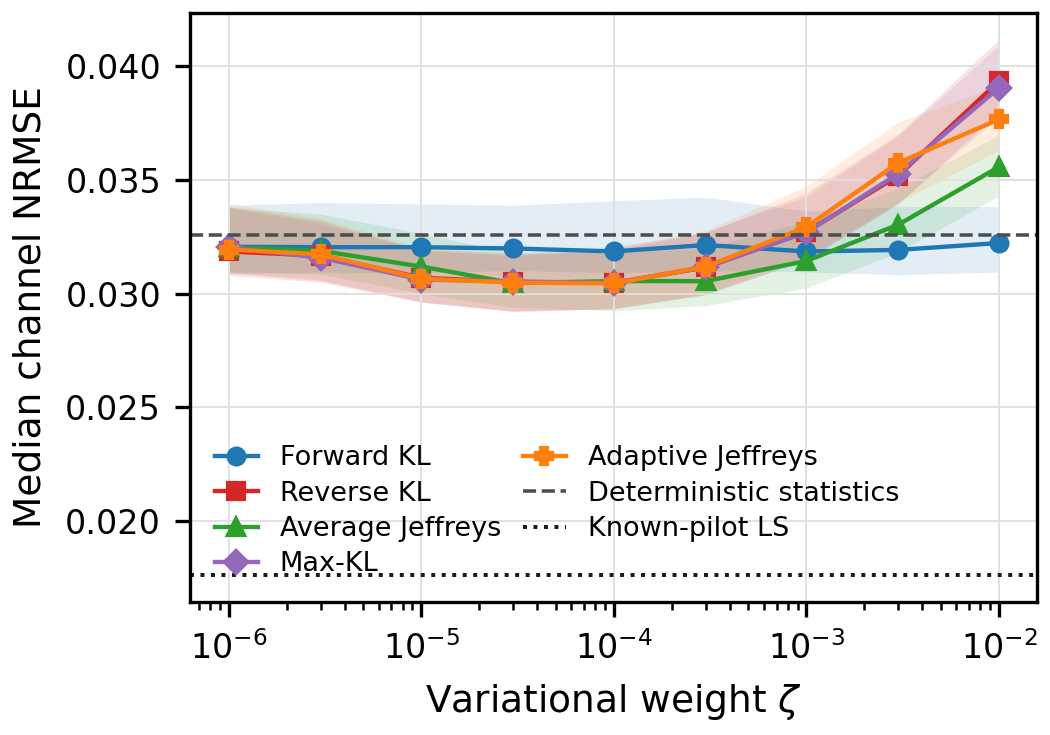}}
    \hfil
    \subfloat[Median normalized KL asymmetry $a_{\mathrm{KL}}$ versus $\zeta$\label{fig:akl_vs_zeta}]
    {\includegraphics[width=0.45\textwidth]{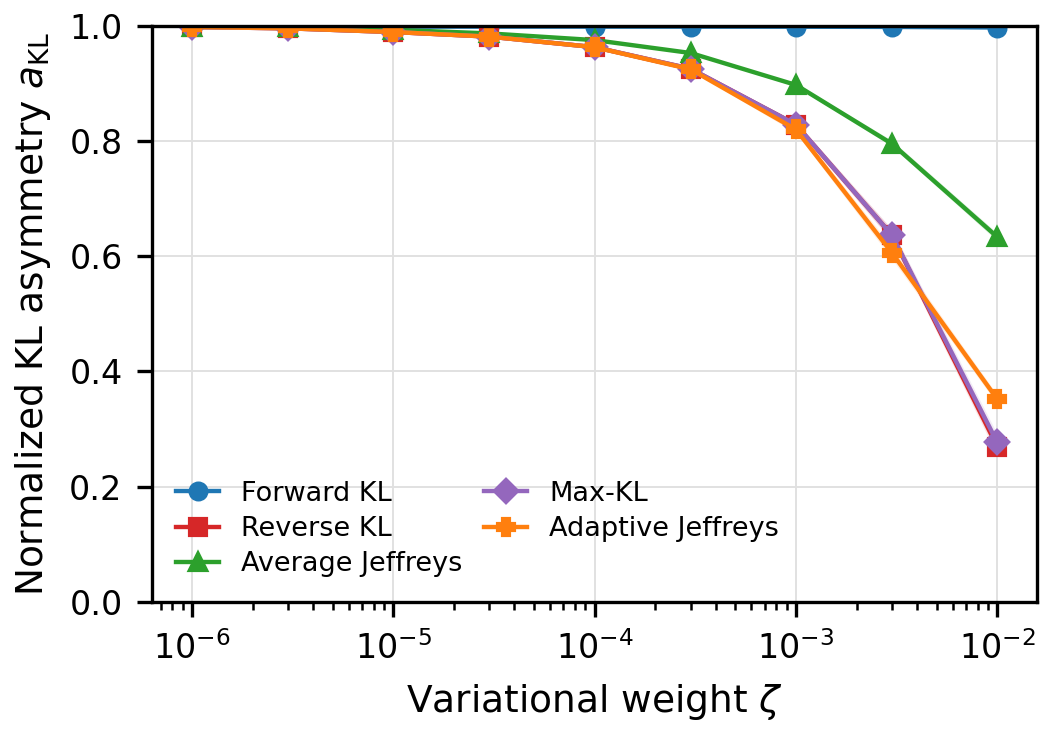}}
    \caption{Effect of the variational regularization weight $\zeta$ on
channel estimation accuracy and normalized directional KL asymmetry over
500 paired Monte Carlo trials. Shaded regions denote the interquartile range.
Reverse KL and the symmetric objectives exhibit a low-error region at
small-to-moderate $\zeta$, while reduced KL asymmetry does not necessarily
correspond to lower channel estimation error.}
    \label{fig:regularization_results}
\end{figure*}
\vspace{-4mm}
\section{Numerical Experiments}
\label{sec:results}

\subsection{Statistics-Based Variational Estimation}

Initial experiments using the raw transmitted and received sample blocks as
neural inputs exhibited poor generalization despite successful optimization.
We therefore replaced the raw representation with the channel-relevant
cross- and auto-statistics
$\bm S_{yx}=\bm Y\bm X^H$ and
$\bm S_{xx}=\bm X\bm X^H$. This representation substantially improved
channel estimation while reducing the neural input to a fixed-dimensional
set of second-order statistics.

Using the statistics-based representation with 1000 training and 1000 test
blocks, the deterministic neural estimator obtained a channel NRMSE of
$0.0323$, while the forward-KL variational estimator obtained an NRMSE of
$0.0311$. For comparison, the known-sample least-squares estimator achieved
an NRMSE of $0.0177$. The variational estimator therefore provides channel
estimates with accuracy comparable to the deterministic learned mapping while
also producing an associated posterior representation. These results motivate the statistics-based representation used in the
subsequent comparison of variational regularization objectives.
\vspace{-3mm}
\subsection{Divergence and Regularization-Weight Sweep}

Fig.~\ref{fig:nrmse_vs_zeta} shows that the conventional forward-KL
objective is nearly insensitive to $\zeta$ over the investigated range,
with median channel NRMSE remaining close to $0.032$. In contrast,
reverse KL and the symmetric objectives exhibit a distinct low-error
region at small-to-moderate $\zeta$, followed by degradation as the
regularization weight increases. At $\zeta=3\times10^{-5}$, Average
Jeffreys reduces the median channel NRMSE by approximately $4.8\%$
relative to the forward-KL VAE at the same regularization weight.

The normalized KL asymmetry in Fig.~\ref{fig:akl_vs_zeta} provides
additional insight into this behavior. Forward KL remains highly
asymmetric across the entire sweep, whereas reverse KL and the
symmetric objectives progressively reduce $a_{\mathrm{KL}}$ as
$\zeta$ increases. However, the strongest reductions in directional
asymmetry occur in the same regime where channel NRMSE begins to
increase. Thus, directional KL balance is not, by itself, an appropriate optimization
target for channel estimation. Excessive
enforcement of directional KL symmetry can coincide with a loss of
channel-state information. The results therefore indicate that both
the form and strength of variational regularization should be selected
according to estimation performance rather than prior conformity or
directional symmetry alone.

Paired analysis across 500 matched trials showed that the relative ordering
of the variational regularizers depended on the regularization weight. At
$\zeta=3\times10^{-5}$, Average Jeffreys reduced channel NRMSE relative to
Forward KL by $0.00210$ on average (paired 95\% CI
$[-0.00240,-0.00181]$), with lower error in 78.4\% of trials. This paired
result supports the corresponding 4.8\% reduction in aggregate median NRMSE.
Near $\zeta=10^{-4}$, Forward KL was inferior to Reverse KL, Average
Jeffreys, Max-KL, and Adaptive Jeffreys, whereas the paired contrasts among
the latter four methods did not resolve a meaningful ranking. The ordering
changed at larger weights. Average Jeffreys yielded lower NRMSE than Reverse
KL, Max-KL, and Adaptive Jeffreys for $\zeta\geq10^{-3}$, while Max-KL and
Adaptive Jeffreys degraded more strongly as the regularization weight
increased. Thus, the paired evidence supports operating-point-specific
conclusions rather than a globally dominant divergence objective.

\begin{figure*}[t!]
    \centering

    \subfloat[Fixed operating-point channel NRMSE\label{fig:exp4_nrmse}]
    {\includegraphics[width=0.46\textwidth] {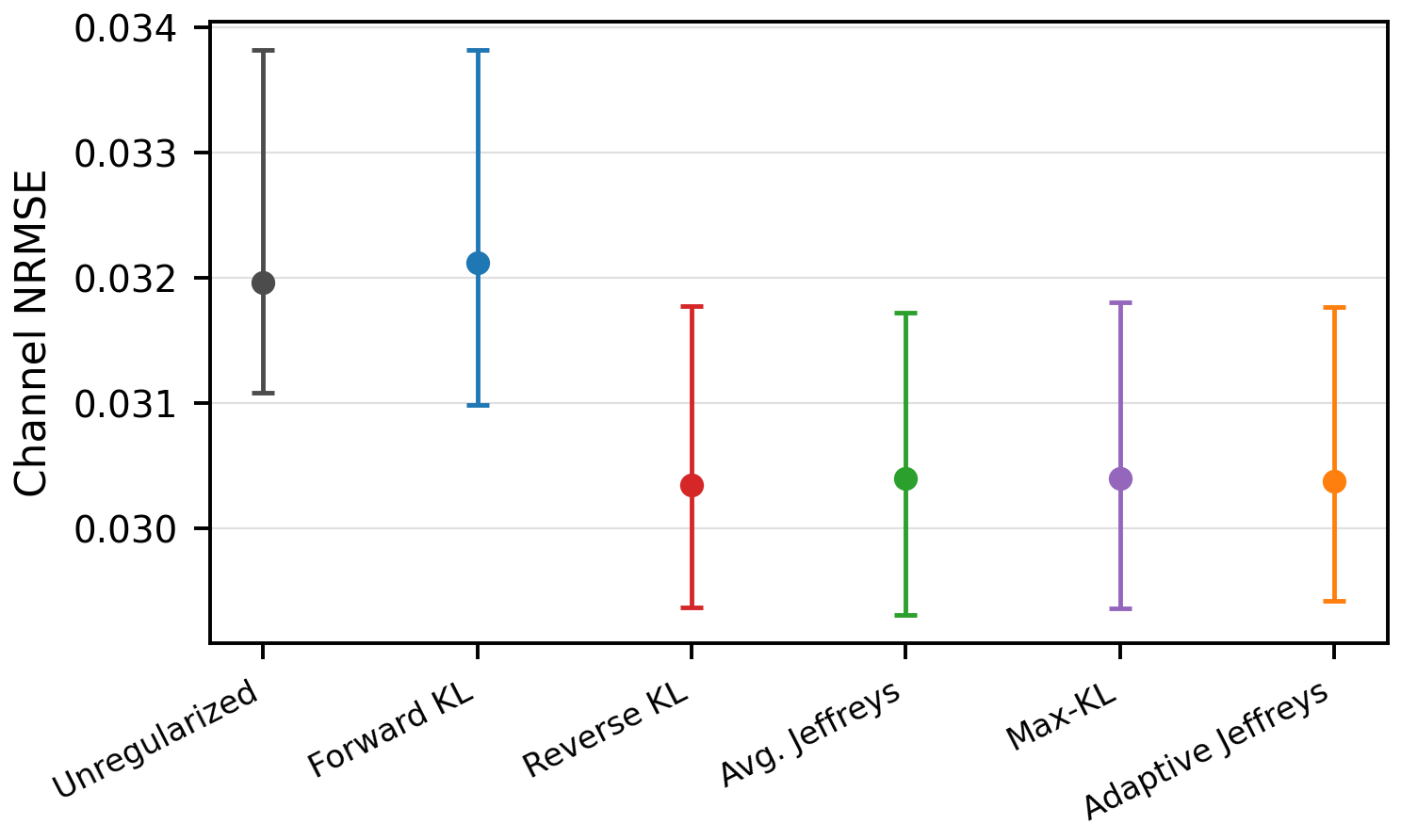}}
    \hfil
    \subfloat[Paired NRMSE difference relative to the unregularized VAE
    \label{fig:exp4_paired}]
    {\includegraphics[width=0.46\textwidth] {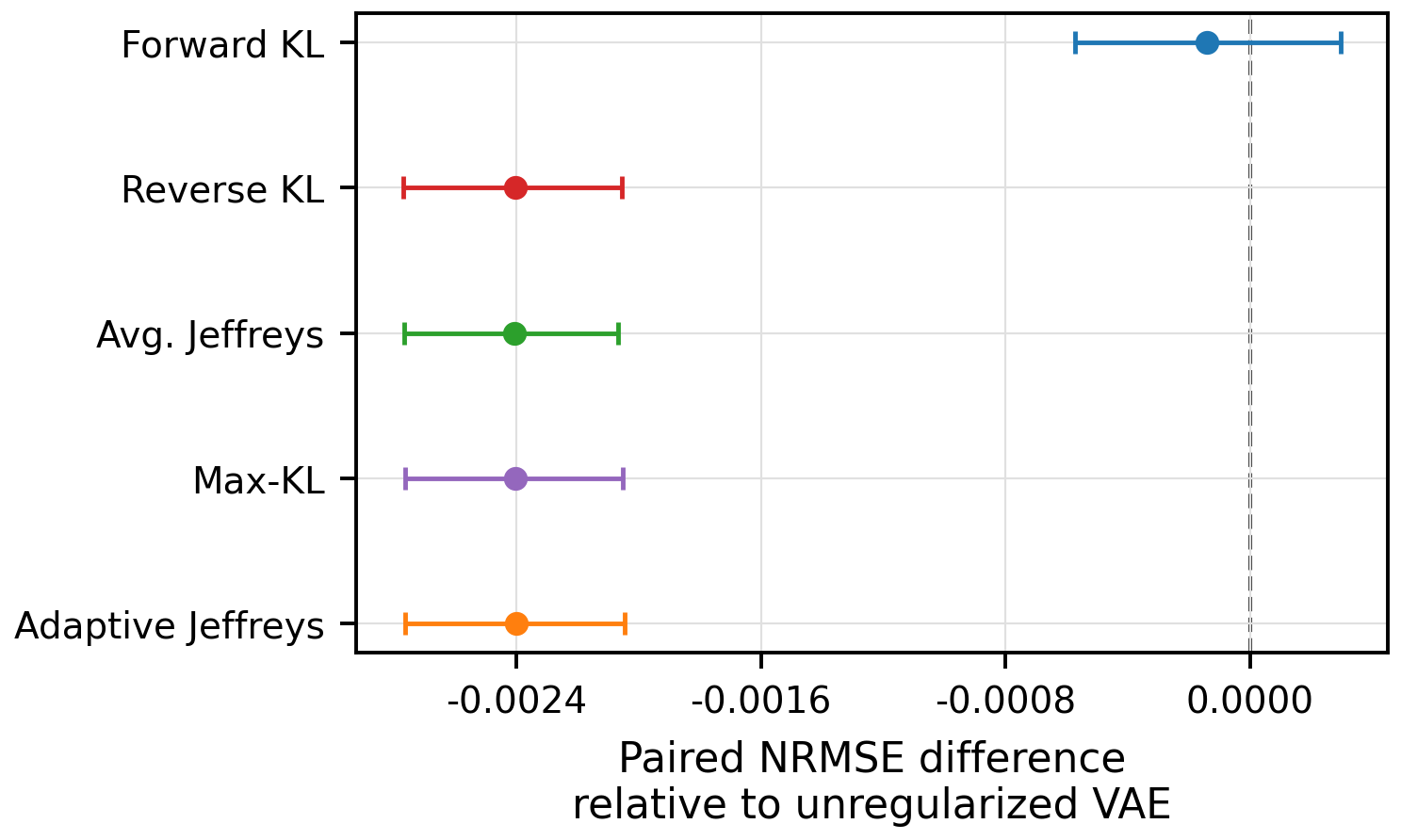}}\\

    \subfloat[Channel NRMSE versus number of samples
    \label{fig:exp5_nrmse}]
    {\includegraphics[width=0.46\textwidth] {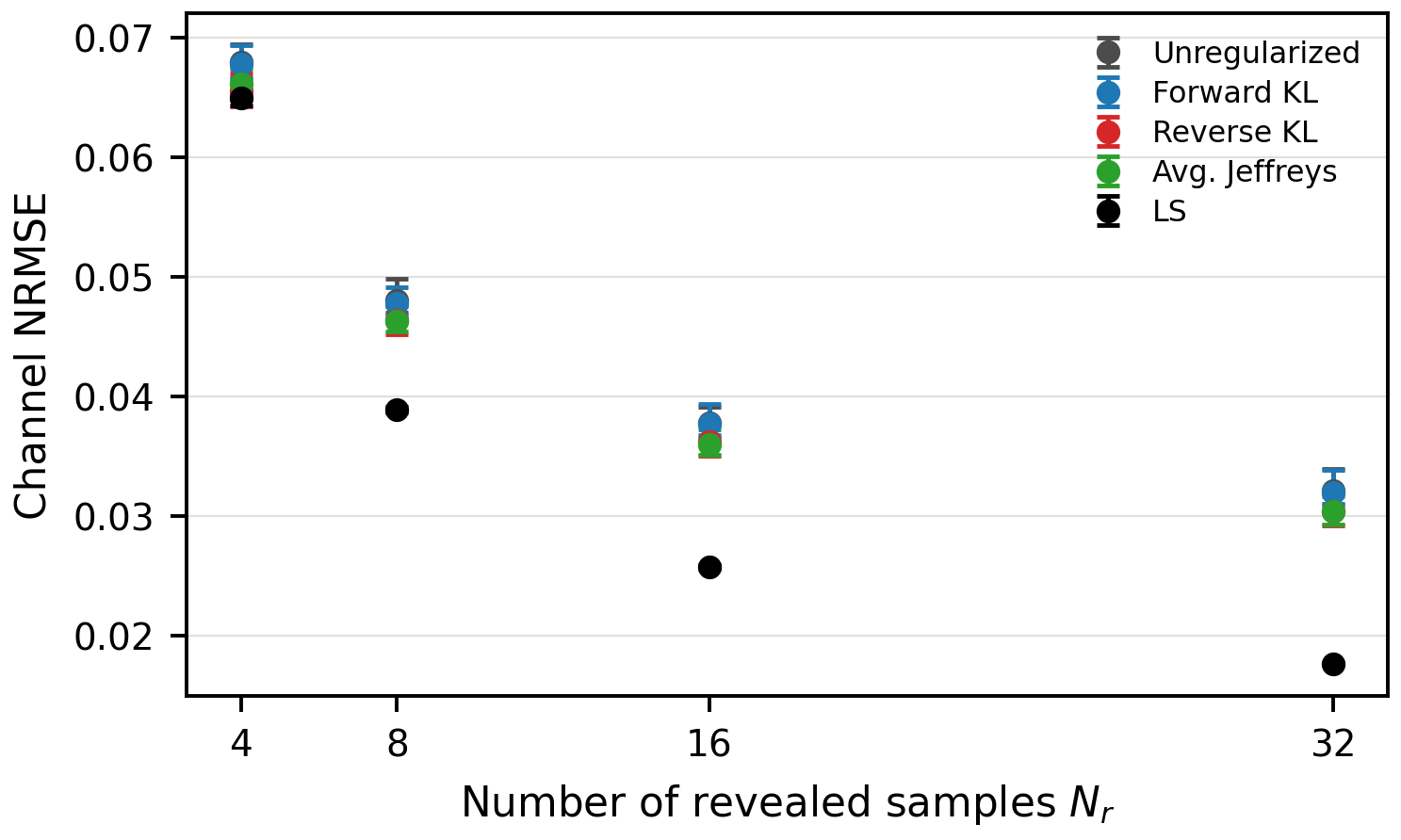}}
    \hfil
    \subfloat[Paired NRMSE difference relative to the unregularized VAE
    versus number of samples
    \label{fig:exp5_paired}]
    {\includegraphics[width=0.46\textwidth] {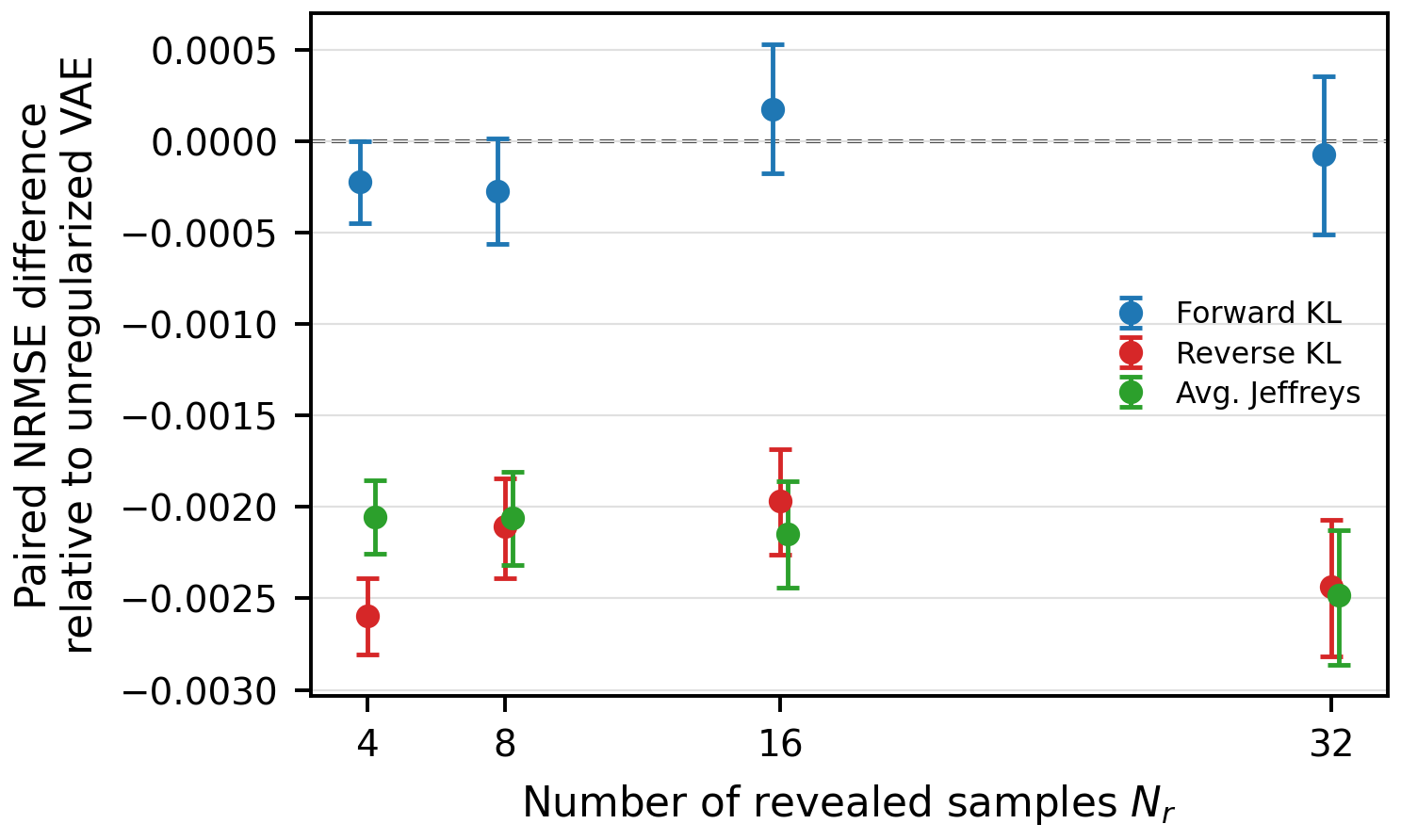}}

    \caption{Regularization performance at fixed operating points and with reduced
sample availability. (a) Median channel NRMSE for the unregularized VAE and selected regularized
objectives, with interquartile ranges. (b) Mean paired NRMSE difference relative to
the unregularized VAE with 95\% paired-bootstrap confidence intervals.
(c) Median channel NRMSE versus the number of samples $N_r$, with
interquartile ranges, for the unregularized VAE, Forward KL, Reverse KL,
Average Jeffreys, and known-sample LS. (d) Mean paired NRMSE difference
relative to the unregularized VAE versus $N_r$ for Forward KL, Reverse KL,
and Average Jeffreys, with 95\% paired-bootstrap confidence intervals.
Negative differences favor the regularized estimator.}
    \label{fig:regularization_control}
\end{figure*}

The sample sweep uses $N_r\in\{32,16,8,4\}$, with the revealed
samples taken as the first $N_r$ columns of each block. For each $N_r$ and
each objective, a new VAE is trained from scratch with a fresh model,
optimizer, and latent-sampling generator. Feature standardization is
recomputed for each $N_r$ from the corresponding 1,000-block training set
and the resulting training-set statistics are applied to the corresponding
test features. The regularization weights are fixed across the sweep at
$\zeta=10^{-3}$ for Forward KL, $\zeta=10^{-4}$ for Reverse KL, and
$\zeta=3\times10^{-5}$ for Average Jeffreys; they are not reselected for
individual values of $N_r$.

\vspace{-3mm}
\subsection{Regularization Control Experiment}

To determine whether the low-error regions observed in the regularization-weight
sweep reflect an effect of variational regularization rather than the stochastic
VAE architecture alone, we performed a confirmatory comparison against the same
estimator trained with $\zeta=0$. The operating point for each regularized
objective was fixed in advance from the preceding sweep and was not retuned
using the control experiment. The fixed values were $\zeta=10^{-3}$ for
Forward KL, $\zeta=10^{-4}$ for Reverse KL, $\zeta=3\times10^{-5}$ for
Average Jeffreys, and $\zeta=10^{-4}$ for both Max-KL and Adaptive Jeffreys.
The 500 confirmation trials were independently generated using root seed
$20260821$, distinct from the root seed $20260817$ used for the preceding
sweep. No training or test blocks or training randomness were reused between
the sweep and confirmation experiment.

Fig.~\ref{fig:exp4_nrmse} shows the fixed operating-point channel NRMSE. The
unregularized VAE obtained a median channel NRMSE of $0.03196$. Forward KL at
$\zeta=10^{-3}$ obtained a similar median NRMSE of $0.03212$. Reverse KL,
Average Jeffreys, Max-KL, and Adaptive Jeffreys obtained median NRMSE values
of $0.03035$, $0.03039$, $0.03040$, and $0.03037$, respectively.

The paired differences relative to the unregularized VAE are shown in
Fig.~\ref{fig:exp4_paired}. For Forward KL, the paired mean difference was not
resolved from zero (95\% CI $[-0.00057,0.00030]$). For Reverse KL, Average
Jeffreys, Max-KL, and Adaptive Jeffreys, the paired mean differences were all
approximately $-0.0024$, with 95\% confidence intervals entirely below zero.

The reduced-revelation results in Figs.~\ref{fig:exp5_nrmse} and
\ref{fig:exp5_paired} show how channel estimation accuracy changes with the
number of samples $N_r$. Channel NRMSE increases as $N_r$ is
reduced, while Reverse KL and Average Jeffreys remain below the
unregularized VAE over much of the sweep. The paired differences in
Fig.~\ref{fig:exp5_paired} show that the magnitude and statistical resolution
of these differences depend on both $N_r$ and the regularization objective.

The confirmatory experiment therefore provides direct evidence that the
low-error operating regions identified in the preceding sweep are not
explained by the stochastic latent-variable architecture alone. At the fixed
operating points, Reverse KL, Average Jeffreys, Max-KL, and Adaptive Jeffreys
each provide a resolved reduction in channel estimation error relative to the
unregularized VAE. Forward KL does not show a resolved improvement. The
control experiment does not establish a meaningful ranking among the four
non-forward objectives.

\vspace{-5mm}

\begin{figure*}[t!]
    \centering
    \subfloat[Median payload NMSE versus number of samples
    \label{fig:exp6_payload_nmse}]
    {\includegraphics[width=0.45\textwidth]{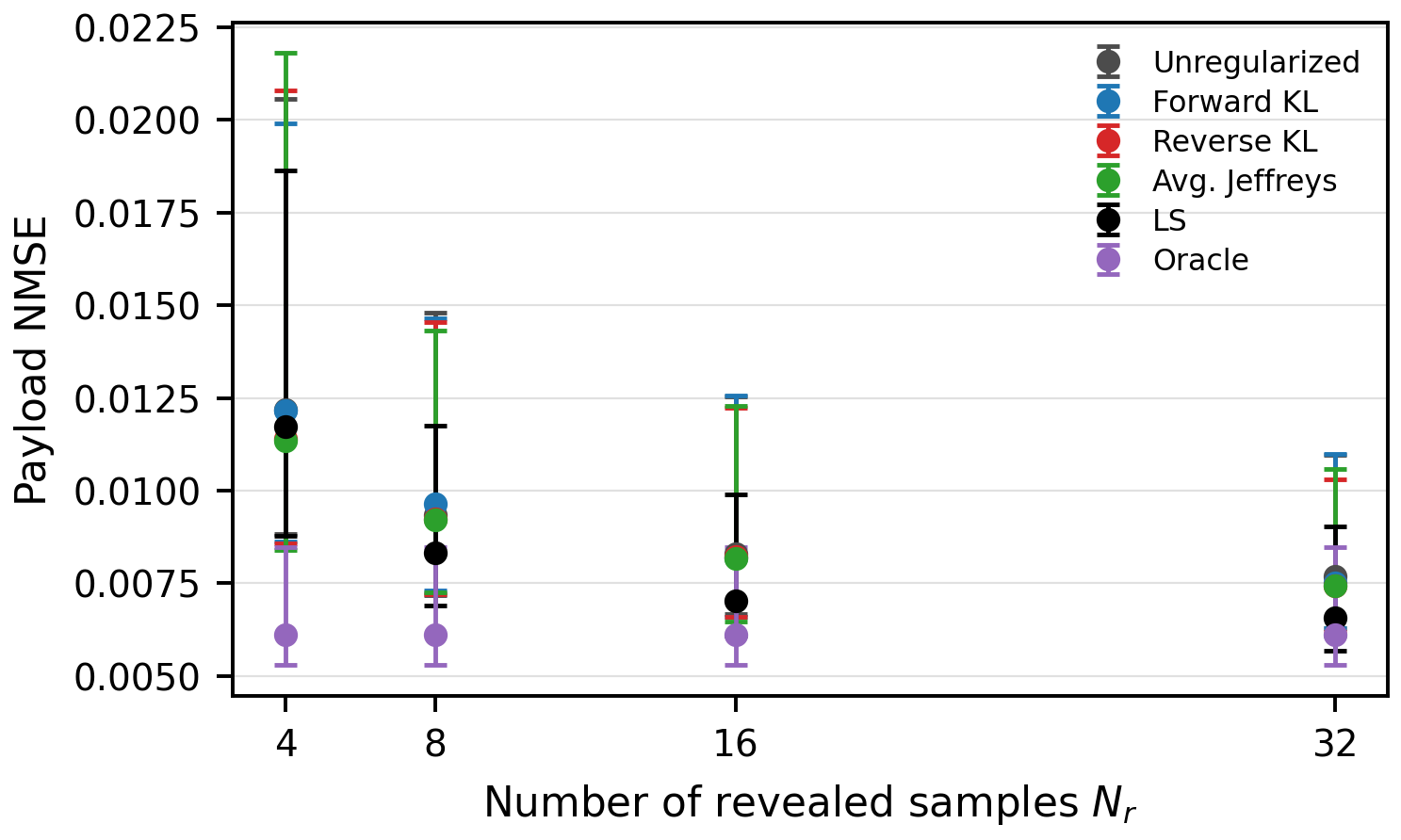}}
    \hfil
    \subfloat[Paired payload-NMSE difference versus number of samples
    \label{fig:exp6_paired_payload}]
    {\includegraphics[width=0.45\textwidth]{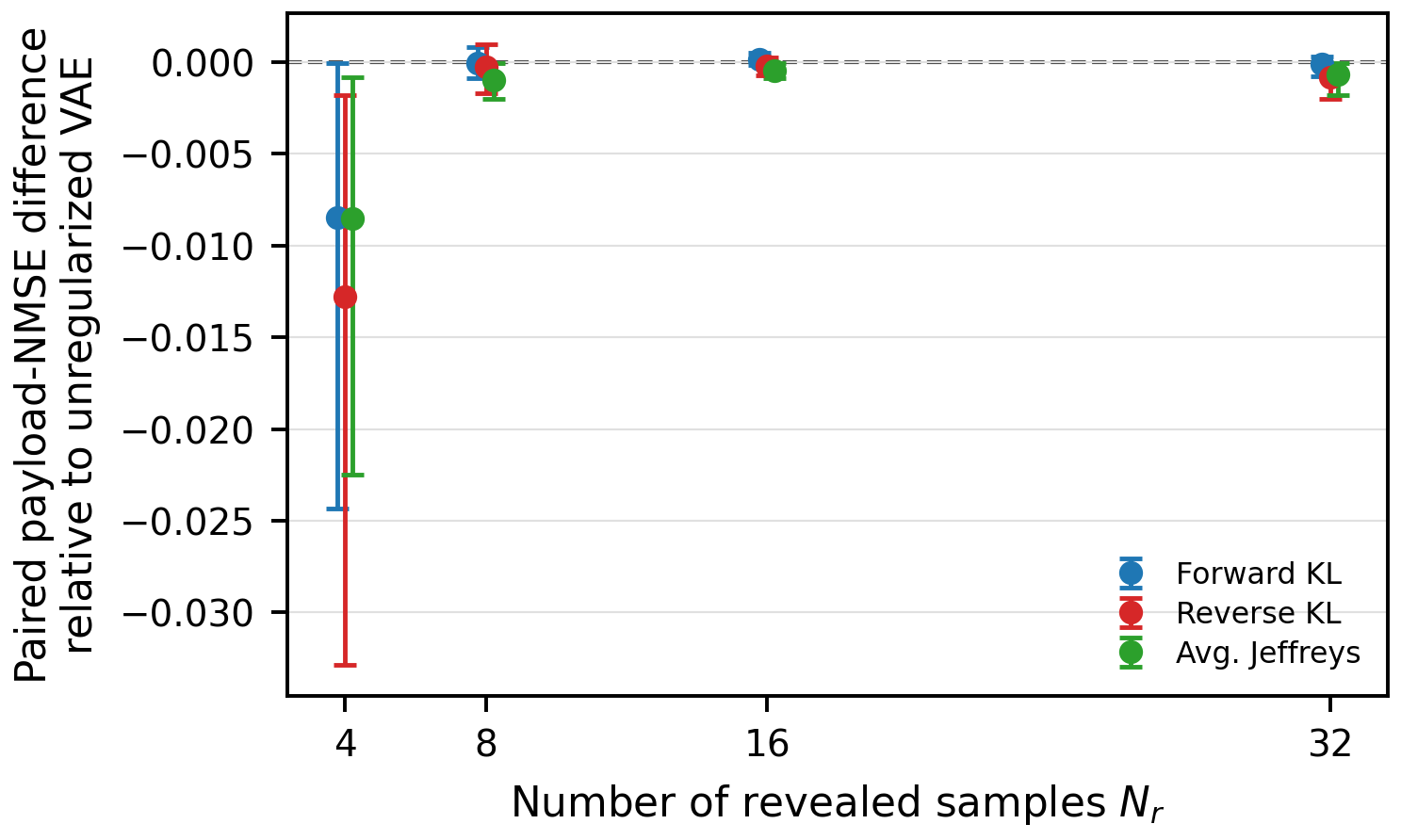}}\\
    \subfloat[Aligned block-NMSE\label{fig:blind_nmse}]
    {\includegraphics[width=0.45\textwidth]{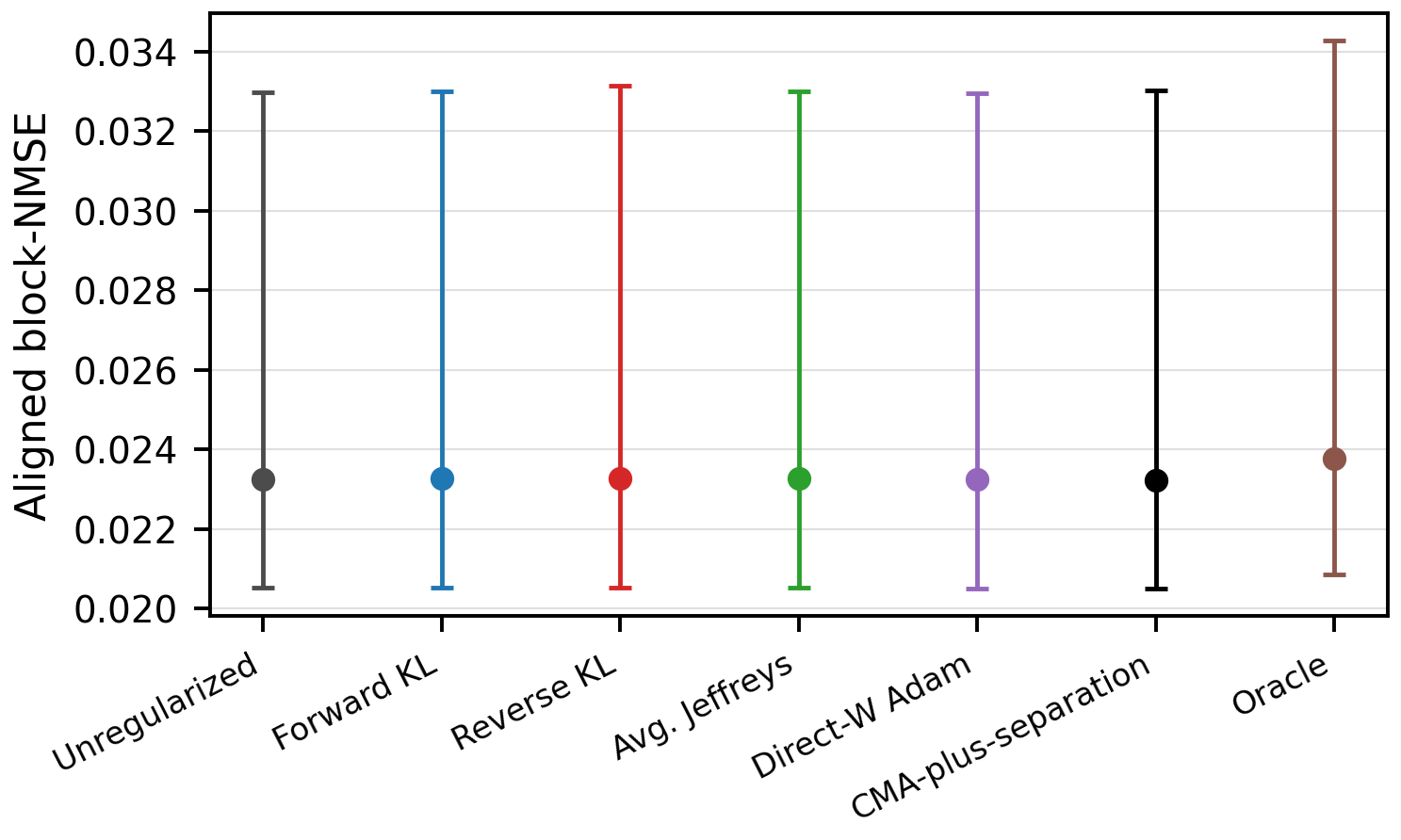}}
    \hfil
    \subfloat[Paired difference relative to CMA-plus-separation\label{fig:blind_cma_paired}]
    {\includegraphics[width=0.45\textwidth]{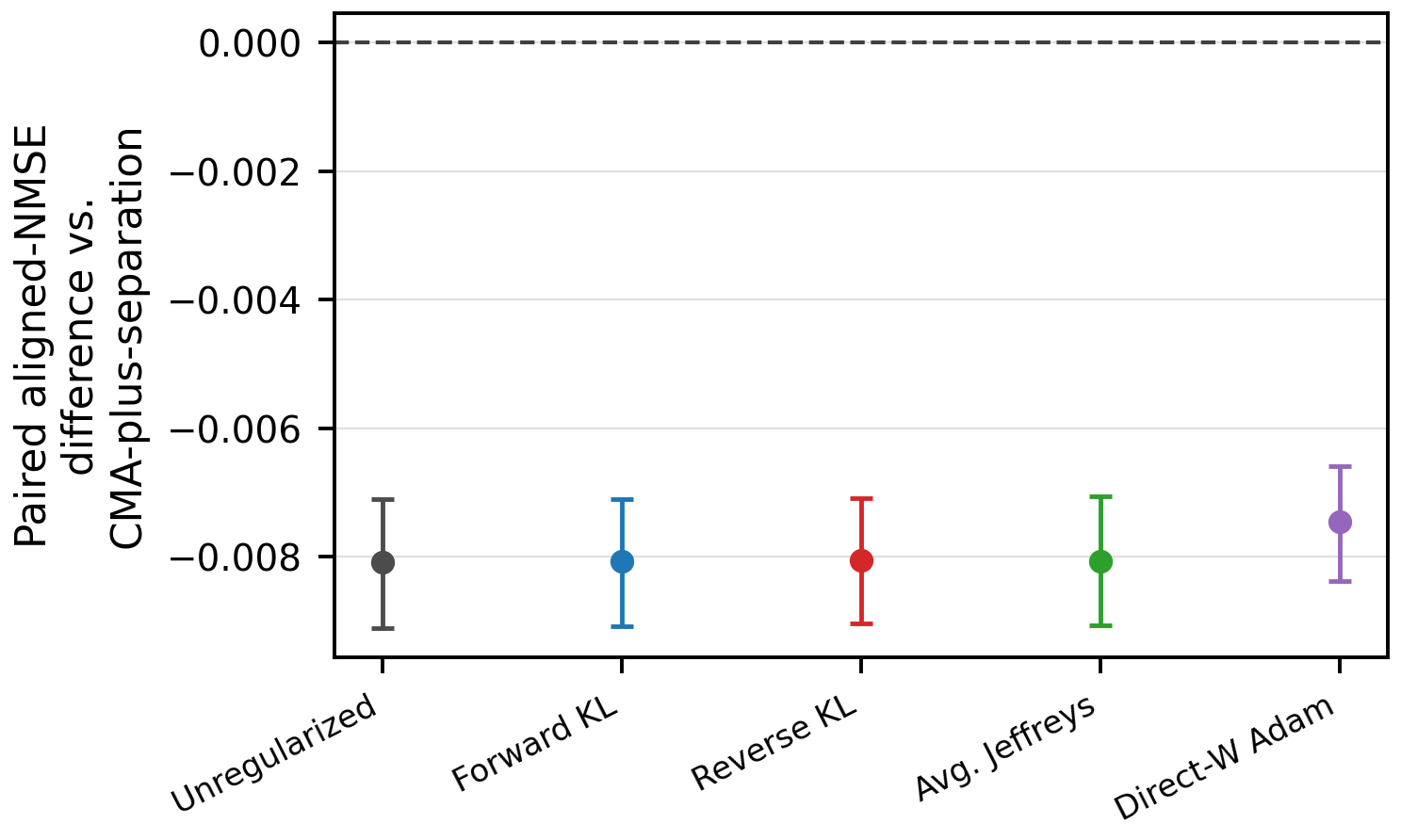}}

    \caption{Receiver-level equalization performance in the semi-blind and fully
blind regimes. (a) Median held-out payload NMSE versus number of samples $N_r$
for the semi-blind estimators, with interquartile ranges. (b) Mean paired
payload-NMSE difference relative to the unregularized VAE with 95\%
paired-bootstrap confidence intervals. Negative values favor regularization.
(c) Median fully blind aligned block-NMSE after received-only carrier-phase
recovery for unregularized VISTA, Forward KL, Reverse KL, Average Jeffreys,
Direct-W Adam, the CMA-plus-separation baseline, and the oracle receiver, with
interquartile ranges. (d) Mean paired aligned-NMSE difference relative to the
CMA-plus-separation baseline with 95\% paired-bootstrap confidence intervals.
Negative values indicate lower aligned block-NMSE than CMA-plus-separation.}
    \vspace{-4mm}
    \label{fig:experiment6_equalization}
\end{figure*}

\subsection{Held-Out Equalization Performance}
To determine whether the channel estimation gains observed under reduced sample availability translate into downstream receiver performance,
we evaluated equalization on a separate hidden payload block that was not
provided to the channel estimator. Fig.~\ref{fig:experiment6_equalization}
shows the resulting held-out payload NMSE and the corresponding paired
differences relative to the unregularized VAE.

The hidden payload comprises $N_u=1000$ samples per polarization,
is generated independently of the revealed block under the same channel
realization, and is not used for channel estimation. Its symbols and
additive receiver disturbance are proper complex Gaussian with variances
$4$ and $0.02$ per complex component, respectively.
For each channel estimate $\widehat{\bm H}$, the payload is recovered by
zero-forcing equalization,
\begin{equation}
    \widehat{\bm X}_u
    =
    \widehat{\bm H}^{\dagger}\bm Y_u,
\end{equation}
where $(\cdot)^\dagger$ denotes the Moore--Penrose pseudoinverse. The
payload NMSE is computed as
\begin{equation}
    \operatorname{NMSE}_{\mathrm{payload}}
    =
    \frac{\left\|\widehat{\bm X}_u-\bm X_u\right\|_F^2}
    {\left\|\bm X_u\right\|_F^2+\epsilon}.
\end{equation}
The oracle receiver uses the true channel in the same zero-forcing operation,
$\widehat{\bm X}_{u,\mathrm{oracle}}=\bm H^\dagger\bm Y_u$, and therefore
retains nonzero error due to receiver noise and zero-forcing noise
enhancement.

As shown in Fig.~\ref{fig:experiment6_equalization}a, payload NMSE increases
as the number of samples $N_r$ is reduced for all non-oracle
receivers. Reverse KL and Average Jeffreys generally achieve lower median
payload NMSE than the unregularized VAE, although the magnitude of the
difference depends on the sample count and regularization objective.
LS remains competitive at the larger sample counts but degrades at $N_r=4$, while the oracle channel provides the expected lower
reference level.

The paired analysis in Fig.~\ref{fig:experiment6_equalization}b shows that
the equalization benefit depends on both $N_r$ and the regularization
objective. Average Jeffreys yields a negative paired mean difference relative
to the unregularized VAE at all four sample counts, with the
95\% paired-bootstrap confidence interval excluding zero in each case.
Reverse KL shows a resolved reduction at $N_r=32$ and $N_r=4$, whereas its
intervals include zero at $N_r=16$ and $N_r=8$. Forward KL provides little
consistent improvement, except for a small resolved reduction at $N_r=4$.

These results indicate that improved channel estimation can translate into
improved recovery of previously unrevealed transmitted samples. The downstream effect depends on $N_r$, the regularization objective, and the
error statistic, so improvements in channel NRMSE do not translate uniformly
into lower payload NMSE.

\vspace{-3mm}
\subsection{Fully Blind Direct-Equalizer Results}
\label{sec:blind_results}

Figs.~\ref{fig:blind_nmse} and~\ref{fig:blind_cma_paired} summarize the fully
blind results using $3000$ paired trials, $1000$ training epochs, and
$1000$ received samples. All blind receivers use received-only fourth-power
carrier-phase recovery before the final discrete QPSK phase and permutation
alignment used for evaluation. The oracle-channel zero-forcing reference uses the true channel matrix,
$\widehat{\bm X}_{\mathrm{oracle}}=\bm H^\dagger\bm Y$. It is included only
as a reference and is not available to the blind receivers. The blind regularization comparison includes Forward KL, Reverse KL, and
Average Jeffreys. Max-KL and Adaptive Jeffreys are not included in this
confirmation experiment.

The four VISTA variants and Direct-W Adam have nearly identical median aligned
block-NMSE values, ranging from $0.02323$ to $0.02326$. In contrast, their
mean aligned block-NMSE values are lower than that of the CMA-plus-separation
baseline. Unregularized VISTA obtains a mean of $0.03388$, compared with
$0.04196$ for CMA-plus-separation, with paired mean difference $-0.00808$ and
$95\%$ confidence interval $[-0.00911,-0.00711]$. Direct-W Adam gives a mean
of $0.03450$, with paired mean difference $-0.00746$ and
$95\%$ confidence interval $[-0.00838,-0.00659]$. Thus, the mean difference
is concentrated in the upper tail rather than in a shift of the central
performance distribution. On the difficult trials, CMA exhibits substantially
larger errors and poorer separation than the learned receivers. The
continuous-phase-aligned diagnostic remains nearly identical to the reported
carrier-recovered NMSE.

Relative to unregularized blind VISTA, Forward KL gives a paired mean
difference of $1.89\times10^{-5}$ with 95\% confidence interval
$[-1.46\times10^{-6},\,4.40\times10^{-5}]$, while Reverse KL and Average
Jeffreys give $2.78\times10^{-5}$ $[5.36\times10^{-6},\,5.61\times10^{-5}]$
and $1.42\times10^{-5}$ $[3.41\times10^{-7},\,2.64\times10^{-5}]$,
respectively. None of the three variational regularizers produces a practically meaningful
equalization improvement over unregularized blind VISTA at the selected
operating points. The matched Direct-W Adam result reproduces most of the
mean reduction relative to the CMA-plus-separation baseline, indicating that
the observed blind-equalization improvement is not dependent on the
variational regularization.

\vspace{-3mm}
\section{Discussion}
\label{sec:disc}
\subsection{Estimation Accuracy and Posterior Regularization}

Variational regularization affects channel estimation through both the
divergence form and its weight. Forward KL remains comparatively insensitive
to $\zeta$, whereas Reverse KL and the symmetric objectives exhibit a
low-error region at moderate $\zeta$, followed by degradation under stronger
regularization, indicating that excessive prior conformity can suppress
information needed for physical channel estimation. The KL-asymmetry results
show that reduced directional imbalance is not sufficient for improved
estimation: several objectives decrease $a_{\mathrm{KL}}$ as $\zeta$
increases even when channel NRMSE increases. Directional KL balance should
therefore be treated as a diagnostic property rather than an estimation
objective. The paired Monte Carlo results reinforce that the relative ordering
of regularizers changes with $\zeta$, with no single divergence uniformly
superior across the investigated range.

For the fully blind receiver, we performed a focused diagnostic of the
diagonal Gaussian posterior using Forward KL, Reverse KL, and Average
Jeffreys at fixed $\zeta$ values of
$3\times10^{-6}$, $10^{-7}$, and $3\times10^{-6}$, respectively, over three
matched trials. We evaluated the final posterior log-variances under lower
clipping bounds of $-8$, $-6$, and $-4$, with upper bound $5$. Under the
implemented $[-8,5]$ interval, no final latent log-variance component reached
either bound. Across the three objectives and three trials, this corresponds
to $0/72$ components at either bound. At the tighter $[-4,5]$ interval,
$3/72$ final components reached the lower bound. The diagnostic monitored
clipping only at final evaluation and does not establish whether clipping
occurred during training.
\vspace{-3mm}
\subsection{Limitations and Security Relevance}

The proposed estimator is motivated by receiver-side parameter estimation in dual-polarization free-space quantum
communication links. Quantities such as the effective channel
matrix, residual error statistics, and associated uncertainty can
enter subsequent calibration and security-relevant processing.
Improved channel estimation, however, is not equivalent to a
CV-QKD security proof.

The reported results therefore quantify estimator-level performance rather than certified protocol security. In particular,
reduced channel NRMSE or improved posterior calibration
does not by itself imply an increase in secret-key rate or
preservation of composable security. Such claims require a
separate protocol-specific analysis that propagates channel
estimation error, receiver assumptions, and uncertainty into the
corresponding security model.

The present study includes channel estimation experiments
and a fully blind QPSK direct-equalizer experiment. The blind
results establish equalizer inference only up to the finite phase permutation ambiguity of the source constellation and do not
establish unique physical-channel recovery. They also use a
static channel and constant-modulus signaling. The results
should therefore be interpreted as receiver-level evidence under
the adopted signal models, not as a complete treatment of
blind operation under general modulation, dynamic free-space
channels, or free-space CV-QKD security.
\vspace{-2mm}
\subsection{Complexity and Implementation}
\label{ss:complexity}

Formation of the cross- and auto-statistics costs $O(N_p)$ per estimation
block, and the resulting 16 real features provide a fixed-dimensional input
whose size is independent of $N_p$. Neural inference is dominated by the
encoder and decoder forward passes; the different KL objectives add
negligible arithmetic cost relative to the network operations.
In the fully blind receiver, the features are formed once per block.
Each VISTA adaptation epoch evaluates the encoder and decoder on the
block-level feature vector and computes the equalization loss over all
$N_b$ received samples. Direct-$W$ Adam and CMA-plus-separation optimize
only the eight real equalizer parameters and avoid this neural-network
overhead.  Least squares
forms the same statistics and inverts the $2\times2$ matrix $\bm S_{xx}$,
so its cost is dominated by statistics formation. The present implementation
evaluates estimator feasibility and equalization accuracy rather than
optimized low-complexity real-time adaptation.

\vspace{-2mm}
\section{Conclusion}
\label{sec:conclusion}

This paper investigated variational regularization for statistics-based
dual-polarization channel estimation and received-only QPSK equalization.
In the revealed-sample setting, Reverse KL and symmetric penalties improved
channel-estimation accuracy over the unregularized VAE at selected operating
points. Stronger regularization could reduce directional KL asymmetry while
degrading estimation accuracy, showing that KL balance is a diagnostic rather
than a performance surrogate.

In the fully blind benchmark, after received-only carrier-phase recovery,
VISTA and Direct-$W$ Adam achieved nearly identical median aligned
block-NMSE. Their lower mean errors relative to the CMA-plus-separation
baseline were concentrated in difficult trials. Direct-$W$ Adam reproduced
most of VISTA's mean improvement, and the tested weak regularizers provided
no resolved benefit over unregularized VISTA.

These findings distinguish the benefits of variational regularization in
revealed-sample channel estimation from the optimization effects observed
in blind equalization. The results concern controlled static receiver
models motivated by free-space quantum communications; dynamic-channel
operation and protocol-level CV-QKD security remain outside scope.

\bibliographystyle{IEEEtran}
\bibliography{IEEEabrv,vista}

\end{document}